%% file: Horse_mnrasv4.tex
\documentclass[useAMS,usenatbib]{mn2e}

\def\Zsol{\hbox{Z$_{\odot}$}}
\def\Msol{\hbox{M$_{\odot}$}}

\def\flux{\hbox{ergs\,s$^{-1}$\,cm$^{-2}$\,\AA$^{-1}$}}

\usepackage{epsfig,natbib,url,graphicx,subfig,float}
\usepackage{lscape,amsmath,amssymb}
\usepackage{booktabs,threeparttable}
\usepackage{longtable,deluxetable}
\usepackage{color}
\usepackage{soul} 
\usepackage{fancyvrb}
\usepackage{pifont}
\usepackage{ulem}
\usepackage{pdflscape}

\newcommand{\hii}{H~{\sc ii}}

\newcommand{\heii}{He~{\sc ii}}
\newcommand{\kms}{km\,s$^{-1}$}
\newcommand{\eld}{$N_{\rm e}$}

\newcommand{\elt}{$T_{\rm e}$}

\newcommand{\foiii}{[O~{\sc iii}]}
\newcommand{\sfoiii}{O~{\sc iii}]}

\newcommand{\foii}{[O~{\sc ii}]}
\newcommand{\fsii}{[S~{\sc ii}]}

\newcommand{\fciii}{[C~{\sc iii}]}

\newcommand{\sv}{S\,{\sc v}}

\newcommand{\mgii}{Mg~{\sc ii}}

\newcommand{\civ}{C~{\sc iv}}
\newcommand{\sIii}{Si~{\sc ii}}
\newcommand{\sfniv}{N~{\sc iv}]}
\newcommand{\sfsIiii}{Si~{\sc iii}]}
\newcommand{\sIiv}{Si~{\sc iv}}
\newcommand{\sfciii}{C~{\sc iii}]}

\newcommand{\feii}{Fe~{\sc ii}}

\newcommand{\alii}{Al~{\sc ii}}
\newcommand{\aliii}{Al~{\sc iii}}

\newcommand{\ha}{H$\alpha$}

\newcommand{\hg}{H$\gamma$}

\newcommand{\Lya}{Ly$\alpha$}

\newcommand{\avgdens}{$3.92<\log($\eld/cm$^{-3})<4.36$}

\title[UV Mapping of the Cosmic Horseshoe]{Mapping UV Properties Throughout the Cosmic Horseshoe: Lessons from VLT-MUSE\thanks{Based on observations collected at the European Organisation for Astronomical Research in the Southern Hemisphere under ESO programme(s) 094.B-0771(A).}}
\author[James et al. ]{Bethan L. James$^{1,2,3}$\thanks{E-mail:bjames@stsci.edu}, Matt Auger$^{2}$, Max Pettini$^{2}$, Daniel P. Stark$^{4}$, Vasily Belokurov$^{2}$
\newauthor  \&\ Stefano Carniani$^{3,5}$\\
$^1$Space Telescope Science Institute, 3700 San Martin Drive, MD, 21218\\
$^{2}$Institute of Astronomy, University of Cambridge, Madingley Road, Cambridge, CB3 0HA\\
$^3$Cavendish Laboratory, University of Cambridge, 19 J.J. Thomson Avenue, Cambridge, CB3 0HE, UK\\
$^{4}$Steward Observatory, The University of Arizona, 933 N Cherry Ave, Tucson, AZ, 85721, USA\\
$^{5}$Kavli Institute for Cosmology, University of Cambridge, Madingley Road, Cambridge, CB3 0HA, UK\\
}
\begin{document}

\date{Accepted 2018 Feb 1. Received in original form Nov 2017}

\pagerange{\pageref{firstpage}--\pageref{lastpage}} \pubyear{2017}

\maketitle

\label{firstpage}

\begin{abstract}
We present the first spatially-resolved rest-frame UV study of the gravitationally lensed galaxy, the `Cosmic Horseshoe' (J1148+1930) at $z=2.38$. Our gravitational lens model shows that the system is made up of four star-forming regions, each $\sim$4--8~kpc$^2$  in size, from which we extract four spatially exclusive regional spectra. We study the interstellar and wind absorption lines, along with \sfciii\ doublet emission lines, in each region to investigate any variation in emission/absorption line properties. 
The mapped \sfciii\ emission shows distinct kinematical structure, with velocity offsets of $\sim\pm$50\,\kms\, between regions suggestive of a merging system, and a variation in equivalent width that indicates a change in ionisation parameter and/or metallicity between the regions. Absorption line velocities reveal a range of outflow strengths, with gas outflowing between $-200 \lesssim v$(\kms) $\lesssim -50$ relative to the systemic velocity of that region. Interestingly, the strongest gas outflow appears to emanate from the most diffuse star-forming region.
The star-formation rates remain relatively constant ($\sim$8--16~\Msol\,yr$^{-1}$), mostly due to large uncertainties in reddening estimates. As such, the outflows appear to be `global' rather than `locally' sourced.  
We measure electron densities with a range of $\log$(\eld)$=3.92$--4.36~cm$^{-3}$, and point out that such high densities may be common when measured using the \sfciii\ doublet due to its large critical density. 
Overall, our observations demonstrate that while it is possible to trace variations in large scale gas kinematics, detecting inhomogeneities in physical gas properties and their effects on the outflowing gas may be more difficult.
This study provides important lessons for the spatially-resolved rest-frame UV studies expected with future observatories, such as JWST.

\end{abstract}


\section{Introduction}
The rest-frame UV spectra of star-forming galaxies provide access to a plethora of spectroscopic features that enable insight into a wide range of galaxy properties. The shapes of interstellar absorption lines teach us about the dynamics of the interstellar gas, whereas the depth can reveal both the density and distribution of absorbing gas along the line of sight. Absorption lines originating in the winds of massive stars give clues as to the metal content and age of the stellar population within which the stars lie \citep{Leitherer:2011, Leitherer:2014}, along with the velocity of the wind itself. Emission lines are also present, providing both electron temperature and density diagnostics, indicators of the ionisation state of the gas, and star-formation rates (SFRs). Finally, information from the absorption and emission line features in the rest-frame UV can enable a unique window into the outflowing properties of both the neutral and ionized phases of the gas.

Galactic outflows are a key component for understanding the evolution of galaxies.  By tracing the cyclic flow of baryonic matter in and out of galaxies, via both galactic-scale winds and the accretion of gas from the intergalactic medium (IGM), we gain insight into a wide range of galaxy evolution phenomena.  These powerful flows of interstellar gas, driven by star-formation and/or active galactic nuclei, can be fundamental in both regulating and quenching star-formation in massive galaxies and, as such, constrain the shape of the mass-metallicity relation \citep{Mannucci:2010,Peeples:2011}.  In addition, the metal-enriched gas from supernova ejecta entrains large amounts of cool ISM and is responsible for polluting the IGM with nucleosynthetic products.  However, some of the most basic questions concerning the physics of outflows (e.g. spatial extension, velocity as a function of radius, magnitude as a function of stellar mass or SFR) remain unanswered.

One of the greatest problems in answering such questions is that galactic winds are traced via the blue-shifts of interstellar absorption lines from cool, outflowing gas, seen against a QSO or the backlight of the stellar continuum. As such, absorption line studies have so far been mostly limited to a single sightline through a galaxy or thousands of single sightlines merged together into one spectrum.  Outflows are universally seen within high-$z$ systems, with most galaxies showing blue-shifted absorption by $\sim100-300$~km\,s$^{-1}$ \citep[e.g.,][]{Shapley:2003,Steidel:2010}.  Despite their ubiquity, there appears to be a wide range of properties in the absorption lines formed by outflowing gas.  We are still essentially ignorant as to the underlying reasons for such diversity - differing covering fraction of the ISM, stellar age, metallicity, or combinations of these may be at play.  It should also be remembered that we integrate light over large spatial scales, making it extremely difficult to disentangle such properties and, to-date, only a handful of studies have attempted to measure the extent of outflows at high-$z$ - most of which are statistical \citep[i.e. stacking spectra as a function of impact parameter, e.g.][]{Steidel:2010}. 

Recently, however, the first spatially resolved study of high-$z$ galactic outflows was made by \citet{Bordoloi:2016}, using \mgii\ and \feii\ along four lines of sight towards a star-forming galaxy at $z=1.7$. Here they revealed `locally sourced' outflows where the properties of the outflowing gas are dominated by the properties of the nearest star-forming region. Mapping of such faint spectroscopic features was enabled by the fact that the target was gravitationally lensed, which both magnifies and elongates the source galaxy.  One difficulty with observing the rest-frame UV of gravitationally lensed targets such as this, is that until recently, optical spectroscopic apertures have been limited to either being long-slits or IFUs (integral field units) with small field-of-views and low sensitivity. As such, observations suffer from either only partial coverage of the complete lensed image, flux loss within the slit, or complex configurations of multiple pointings.  Fortunately, these problems can now be overcome with the advent of VLT-MUSE's large 1~arcmin$^2$ FoV and high throughput, finally enabling us to accurately map the UV properties of galaxies at $z=$2--4 in detail with a single pointing \citep[][]{Patricio:2016, Swinbank:2015,Wisotzki:2016,Maseda:2017}.

\begin{figure*}
\includegraphics[scale=0.7]{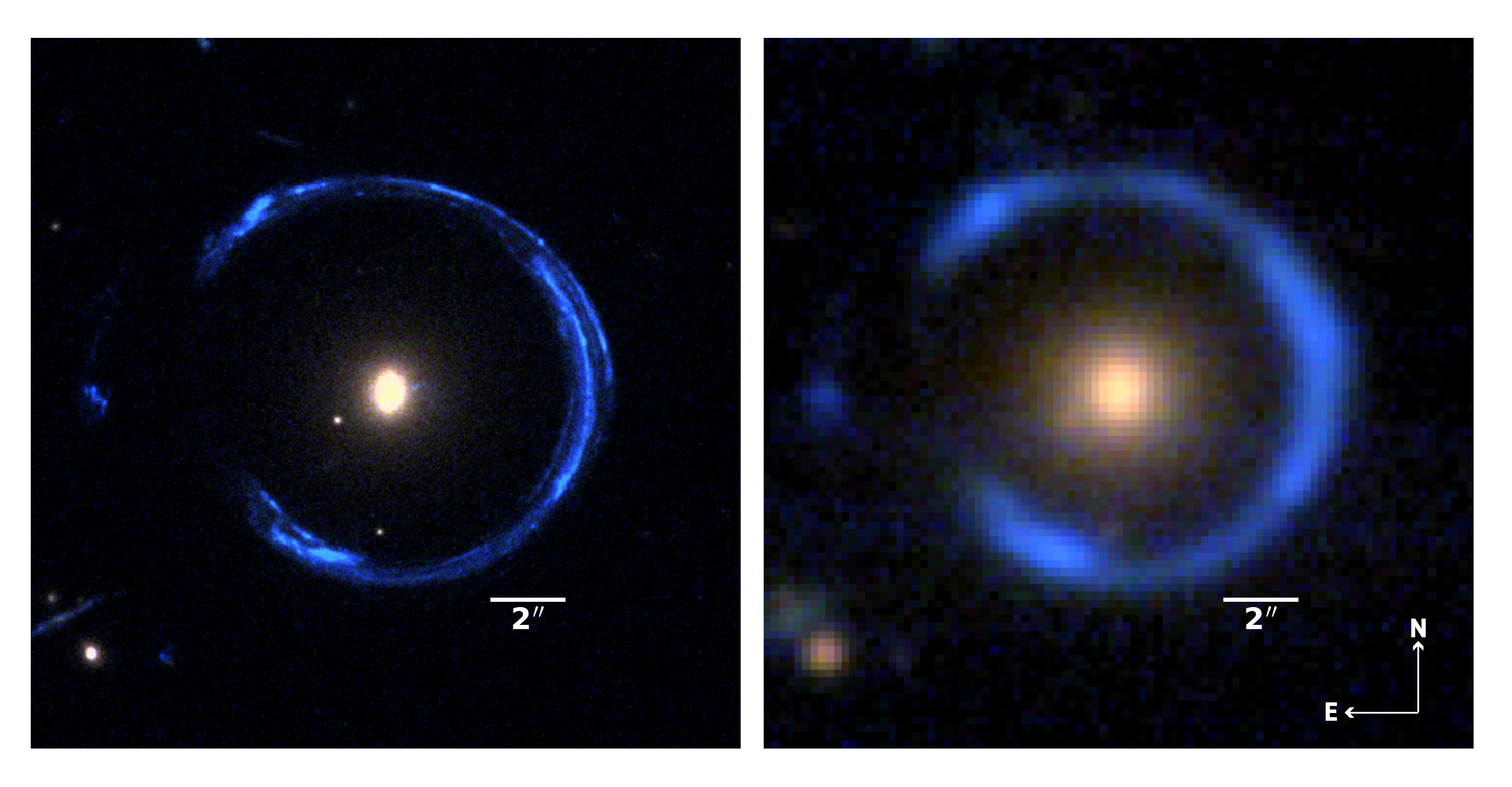}
\caption{Colour composite images of the Cosmic Horseshoe derived from \textit{left:} HST-UVIS images (B=F475W, G=F606W, R=F814W) and \textit{right:} the VLT-MUSE data cube, created from 100~\AA\ channels at 4800~\AA\ (blue), 6050~\AA\ (green), and 8050~\AA\ (red). Both images are $18''\times18''$ in size. All VLT-MUSE observations were made at a seeing of 0.8\arcsec.}\label{fig:colorFig}
\end{figure*}

Here we capitalise on this advancement in instrumentation and present the first spatially-resolved rest-frame UV study of the gravitationally lensed galaxy, the `Cosmic Horseshoe' \citep[J1148+1930][]{Belokurov:2007}.  This galaxy is ideal for a study of this kind for the following reasons: (i) its image is an almost complete Einstein ring (11$''$ diameter, Figure~\ref{fig:colorFig}), fully covered by the MUSE 1~arcmin$^2$ aperture; (ii) it is highly magnified \citep[$\mu\sim24$,][Q09 hereafter]{Quider:2009}, therefore allowing us to probe sub-kpc scales; (iii) its redshift ($z=2.38$) places a suite of absorption lines from different ionisation stages within the MUSE extended wavelength range (465-930~nm), covering \sIiv~$\lambda$1393,1402, \civ~$\lambda$1548, 1550, \sIii~$\lambda$1260, 1526, \alii~$\lambda$1670, and \aliii~$\lambda$1854, 1862 - many of which are known to be strong (Q09); (v) finally, the Cosmic Horseshoe is very well studied and has a large amount of ancillary data, allowing for an optimized mass-model. Alongside a plethora of $HST$ imaging data, the Cosmic Horseshoe has also been partially observed with the infrared IFU `OSIRIS' by \citet{Jones:2013}, whose optical emission line maps reveal variations in \ha\ surface brightness, ordered rotational motion and metallicity variations between the regions.  The observations presented in this study, however, are designed to explore possible variations in properties throughout the Cosmic Horseshoe, as traced by the \textit{rest-frame UV}, including gas outflow velocities, covering fractions, electron density and temperature, star-formation rate, and gas kinematics (e.g. velocity dispersions and kinematics within the system itself).  As such, we aim to shed light on whether outflows are globalized or localized in nature, and if localized, investigate the level to which they are interconnected with the physical properties of the star-forming region from which the gas originates.

The paper is organised as follows. In Section~\ref{sec:data} we document our VLT-MUSE data and its reduction. Section~\ref{sec:model} describes the gravitational lens model and methods used to extract the spatially distinct regional spectra. The absorption line properties (i.e. optical depth, gas kinematics, and equivalent width) are described in Section~\ref{sec:abs}, and emission line properties (i.e. electron density, reddening, star-formation rate, and star-formation rate density) are considered in Section~\ref{sec:emlines}. We discuss and conclude our findings in Section~\ref{sec:disc_conc}.

\section{Observations And Data Processing}\label{sec:data}

\subsection{Observations}
Observations were made using the VLT-MUSE IFU instrument in wide field mode, which delivers medium resolution spectroscopy (R$\sim$1770--3590) at a spatial sampling of 0.2$''$/spaxel (spatial pixel), over a 1$\times$1~arcmin FoV. The extended wavelength range was used, covering 4650--9300 \AA. The orientation of the MUSE FoV is shown in Figure~\ref{fig:colorFig} (zoomed in to the central $18''\times18''$). Separate sky observations were not made due to the small area covered by the target galaxy within the FoV.  Instead, sky spectra were determined from sky-only regions away from the target galaxy. Observations for this program (094.B-0771(A)) were split over four 1~hour observation blocks made between January and April 2015, and another in April 2016, totaling $8\times1500$~s exposures with a total exposure time of 3.3 hours. The original plan consisted of 15 hours of observations, but unfortunately this service program was terminated by ESO before it had been completed. Airmass ranged from 1.39 to 1.47, while seeing remained at 0.8$''$ (at 5000~\AA ) for all OBs.

\subsection{Data Reduction}
The spectroscopic data were reduced using the ESO MUSE standard pipeline (version 1.4) \footnote{\url{https://www.eso.org/sci/software/pipelines/muse/}}. MUSE consists of 24 separate IFUs and the principal aim of the pipeline is to convert the raw data from each CCD into a combined datacube, after correcting for instrumental and atmospheric effects. The pipeline handles all of the standard reduction procedures, including bias, dark, flat-fielding corrections, and illumination corrections, along with geometric corrections, basic sky subtraction, wavelength and relative flux calibrations, and astrometry. Each MUSE spectrum has an associated error array, which is propagated accordingly throughout the reduction pipeline. After reducing each of the eight exposures individually, they were combined into a final single datacube. Flux calibration was performed using standard star observations of GD153, GD108, and EG274.  The final data cube is shown in Figure~\ref{fig:colorFig} as a colour composite made from 100~\AA\ channels at 4800~\AA, 6050~\AA, and 8050~\AA.

\subsection{$HST$ data}

The WFC3 instrument was used to obtain F475W, F606W, and F814W images with the UVIS camera and F110W and F160W images with the NIR camera. These data were reduced using the {\sc astrodrizzle} package, creating final images that are centered on the lensing galaxy and have a pixel scale of 0\farcs04 for the UVIS data and 0\farcs08 for the NIR images. A second image was also created that was individually centered for each filter on a nearby star. This star was used as a model for the point spread function (PSF) during our subsequent modeling. A two-Sersic component model for the lensing galaxy was fitted in all five bands and subtracted from the data. A colour image created from the UVIS data is shown in Figure~\ref{fig:colorFig}.

\section{Lensing Model and Spectral Extraction}~\label{sec:model}
\begin{figure*}
\includegraphics[scale=0.9]{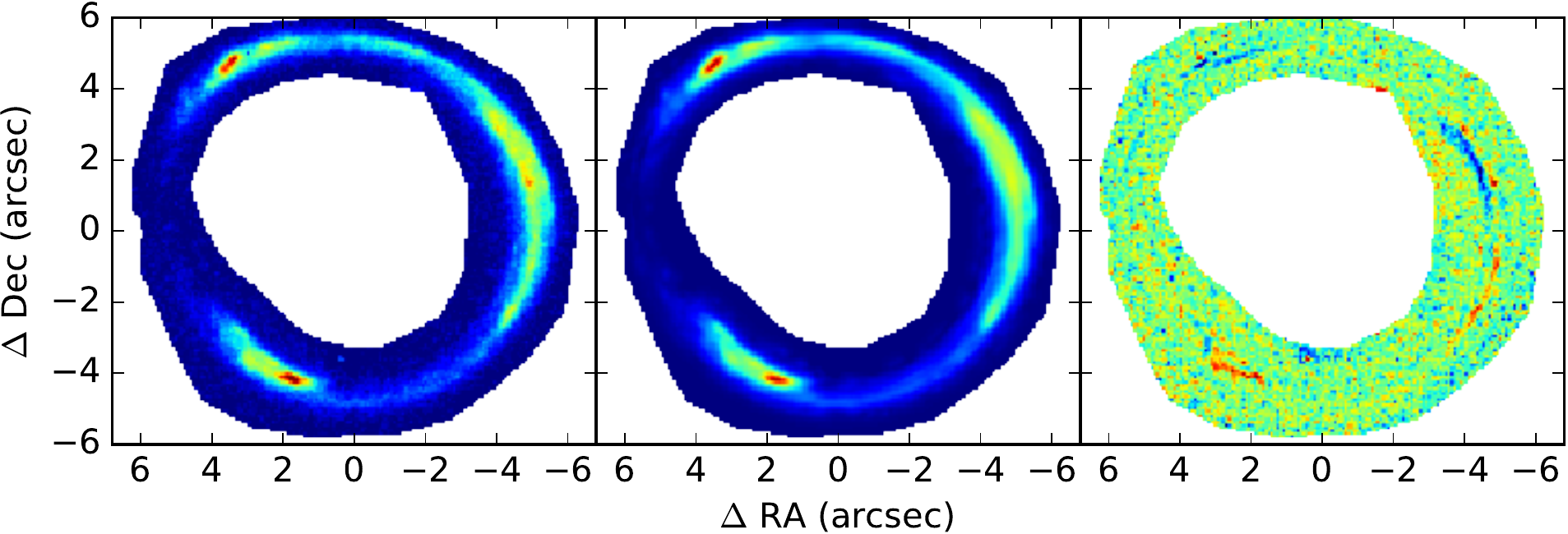}
\caption{Lensing model visualisation for the Cosmic Horseshoe: (left-panel) $HST$ F160W image; (middle-panel) image-plane reconstruction of the Cosmic Horseshoe with the lensing model described in Section~\ref{sec:model}; (right-panel) residuals from subtracting the model from the data.} \label{fig:model}
\end{figure*}

\begin{figure*}
\includegraphics[scale=0.6]{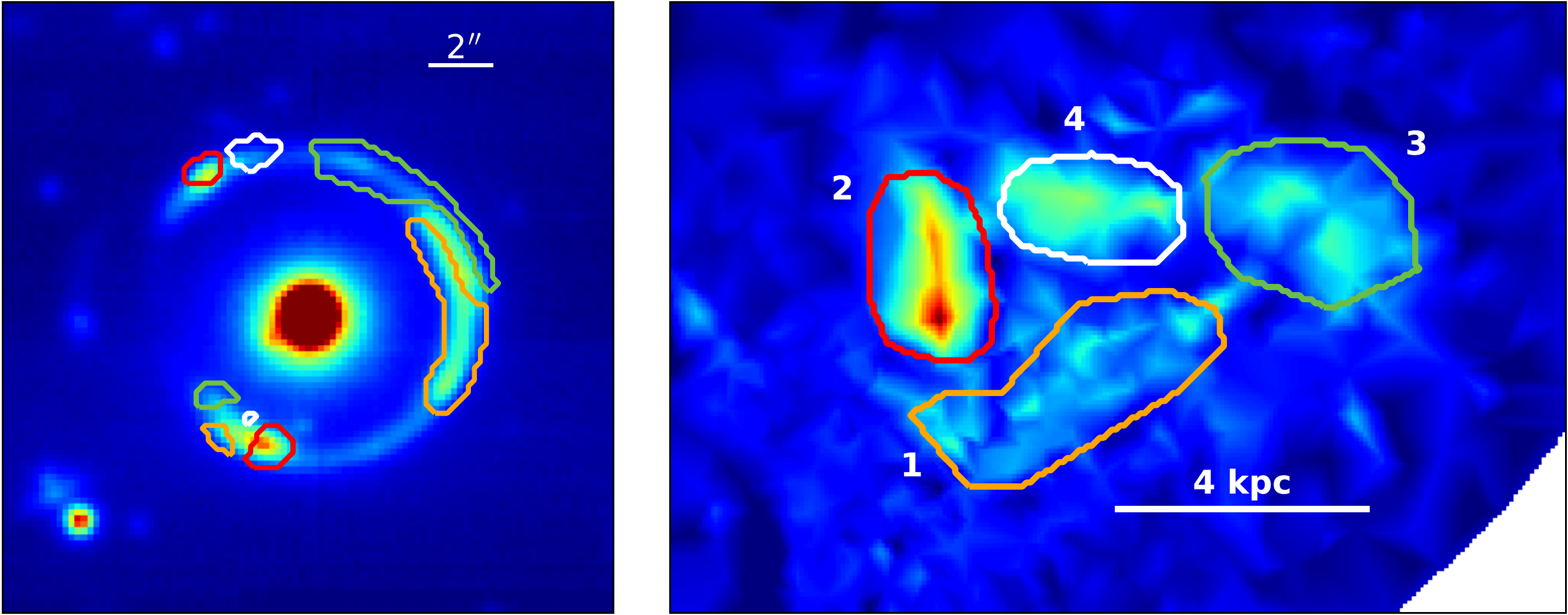}
\caption{Extraction apertures used to define the four separate star-forming regions analysed throughout the paper which we discuss in Section~\ref{sec:model}. Left-panel shows the apertures in the MUSE image plane (continuum region), where apertures were defined in such a way to prevent spatial overlap in the image plane and enable exclusive regional spectra. The right-panel shows the same apertures overlaid in the source-plane reconstruction of the Cosmic Horseshoe ($HST$ F160W) with the lensing model described in Section~\ref{sec:model}. The regions were first defined in the $HST$ source plane and then applied to the MUSE image plane.}\label{fig:apertures}
\end{figure*}

We model the lensing following the procedure developed in \citet{Vegetti:2009} and applied to a set of ten lenses in \citet{Vegetti:2014}. The lensing potential is modeled as an elliptical power-law mass distribution with an external shear component, and the source is modeled on an adaptive grid that approximately follows the magnification distribution of the lens. The model was fitted to the $HST$ F160W data and the best-fit model was subsequently applied to the four other $HST$ images (F110W, F475W, F606W, and F814W) and the MUSE datacube by using the position of the lensing galaxy to align all of the datasets. The F160W data was utilised here because its slightly larger pixel scale (0\farcs08 vs. 0\farcs04) makes it more computationally tractable during modeling. The best-fit model is shown in Figure~\ref{fig:model}, along with the data used to constrain the model and the residuals scaled by the noise in each pixel. We show the best-fit reconstructed source in Figure~\ref{fig:apertures}.

We identify four surface brightness `regions' of the galaxy (shown in Figure~\ref{fig:apertures}) that can be well-separated in the $HST$ source plane and then apply these regions in the MUSE image plane. Region 1 lies within the lensing caustic and is the most highly magnified region with an area-weighted magnification (i.e., not weighted by surface brightness or luminosity) of 22.5$\pm$0.4. This part of the source has a filamentary structure ($\sim$4\,kpc in length) which is seen in all bands. Region 2 consists of the highest surface brightness regions in the source and has a magnification of 12.1$\pm$0.1; this part of the galaxy appears as one elongated structure in the F160W data ($\sim3\times1.7$\,kpc across) but is clearly resolved into two surface brightness peaks in the optical bands. We define this region as the `main star-forming region'. Region 3 also has two clear components, both roughly 1\,kpc in diameter and with fairly low surface brightness, and a magnification of 20.0$\pm$0.1. Finally, Region 4 is the least highly magnified region with a magnification of 5.4$\pm$0.1 but is the second highest surface brightness region in the source and $\sim3\times1.5$\,kpc in size. Note that the uncertainties on the magnifications are the \textit{formal} uncertainties for the powerlaw mass model. However, the residuals indicate that there is additional structure in the mass distribution, and we therefore assume 10\% uncertainties going forward (this is a conservative estimate, as preliminary mass models that include satellites and higher-order angular structure lead to very similar source reconstructions; Auger et al., in prep).  At first sight, this fragmented structure may be somewhat surprising; however, from the UV imaging of Lyman Break Galaxies (LBGs) of \citet{Law:2007} we can see that irregular morphologies are in fact quite typical at $z=$2--3. Our model is not in agreement with \citet{Jones:2013} who suggest that the emission in the Cosmic Horseshoe originates from an elongated region 0.2\,kpc $\times$ 0.4\,kpc, with the full system being $\sim2\times2$~kpc in size. The ($\sim$4 times) difference in scales between the two models may be because \citet{Jones:2013} model assumed quadruple imaging for the two highest surface brightness regions (rather than doubly imaged) which would lead to higher magnification and a smaller intrinsic size in the source plane.

The significant point spread function (PSF) of the spectroscopic data blends parts of the source that are otherwise physically distinguishable. As such, it is necessary to carefully extract spectra from non-overlapping apertures in the MUSE image plane and re-scale them according to their intrinsic fluxes. We first map the entire source back into the image plane and convolve with the MUSE PSF; this essentially produces a synthetic MUSE image. We then map each source region individually, convolve with the PSF, and take the ratio of this `single-component' image to the total flux image. Any pixels where the ratio is greater than 0.75 (0.5 in the case of Region 4) are dominated by a single component, and these pixels constitute the apertures from which we extract each spectrum; these apertures are shown in Figure \ref{fig:apertures}. This procedure clearly excludes significant amounts of the observed MUSE data cube, and to reconstruct the total flux of each region -- and simultaneously account for the lensing magnification -- we normalise the spectra to give the same synthetic WFC3-UVIS F606W flux as is found in each region of the reconstructed source in that filter (F606W fluxes are listed in Table~\ref{tab:regions}, along with $B-V$ and $V-I$ colours).

\begin{table}
\caption{Properties of the four source regions described in Section~\ref{sec:model} and shown in Fig.~\ref{fig:apertures}: magnification factors, physical areas, $HST$ $V$-band de-lensed magnitudes used to scale the spectra extracted from non-overlapping apertures, and photometric colours $B-V$ and $V-I$ from $HST$ imaging. }
\begin{center}
\begin{tabular}{cccccc}
\hline
Region ID & $\mu$ & Area (kpc$^2$) & $V$ (mag) & $B-V$ & $V-I$\\
\hline
1 & 22.5$\pm$0.4 & 8 & 25.21 & 0.208 & 0.103\\
2 & 12.1$\pm$0.1 & 5 & 24.59  & 0.113 & 0.072\\
3 & 20.0$\pm$0.1 & 7 & 25.64 & 0.252 & 0.147\\
4 & 5.4$\pm$0.1 & 4 & 25.75 & 0.211&  0.168\\
\hline
\end{tabular}
\end{center}
\label{tab:regions}
\end{table}%

\begin{figure*}
\includegraphics[scale=0.7]{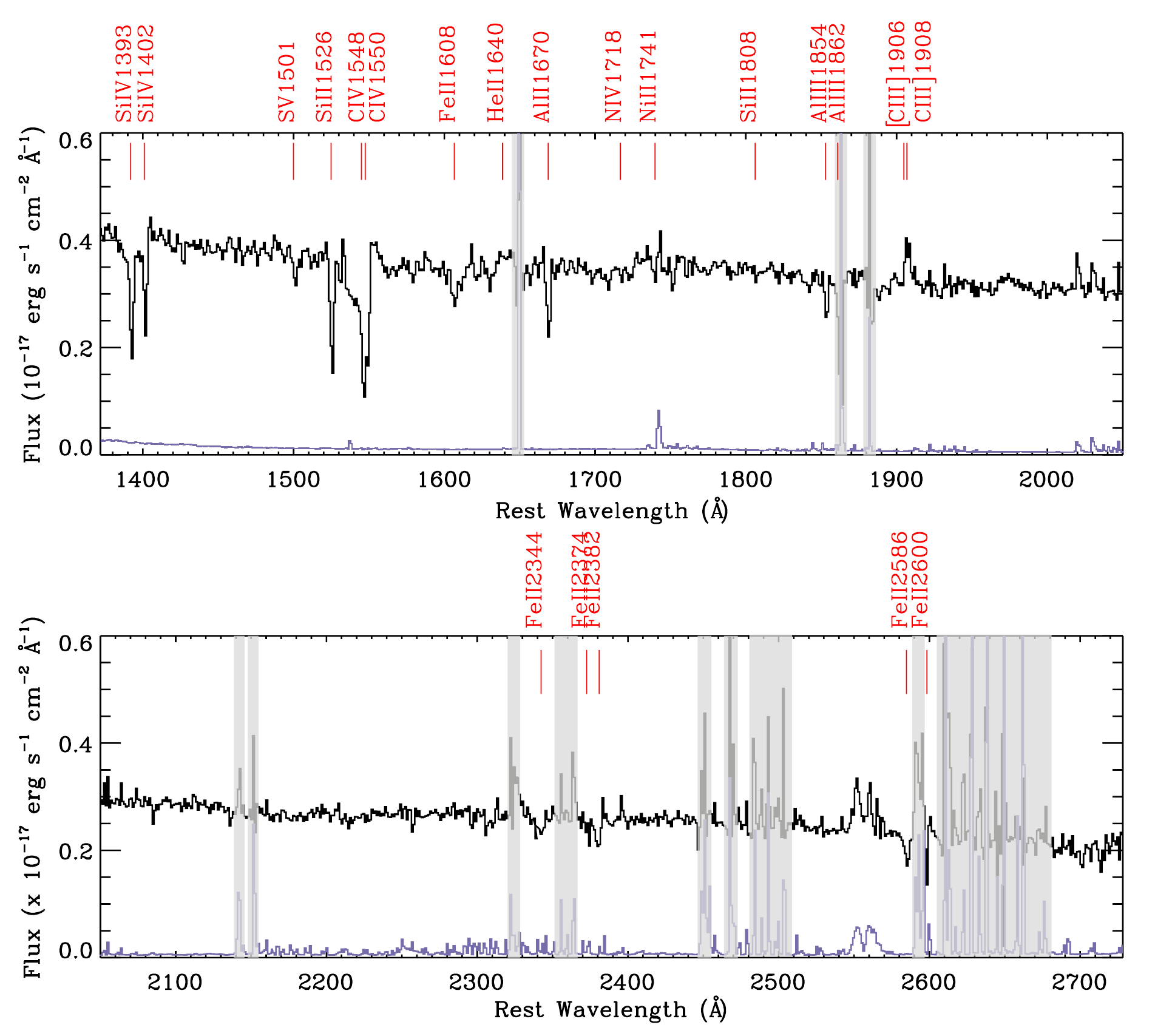}
\caption{MUSE `total' spectrum of the Cosmic Horseshoe, created by summing the four regionally extracted spectra discussed in Section~\ref{sec:model}, with the most important absorption and emission lines labelled. The corresponding `total error' spectrum is shown in blue. Wavelengths are given in the rest frame of the gravitationally lensed galaxy, $z=2.38115$, as measured by \citet{Quider:2009}. Spectra have been smoothed with a 3-pixel boxcar for presentation purposes. Vertical grey boxes indicate regions of elevated noise from skylines.}~\label{fig:spectrum}
\end{figure*}

\begin{figure*}
\includegraphics[angle=90,scale=0.9]{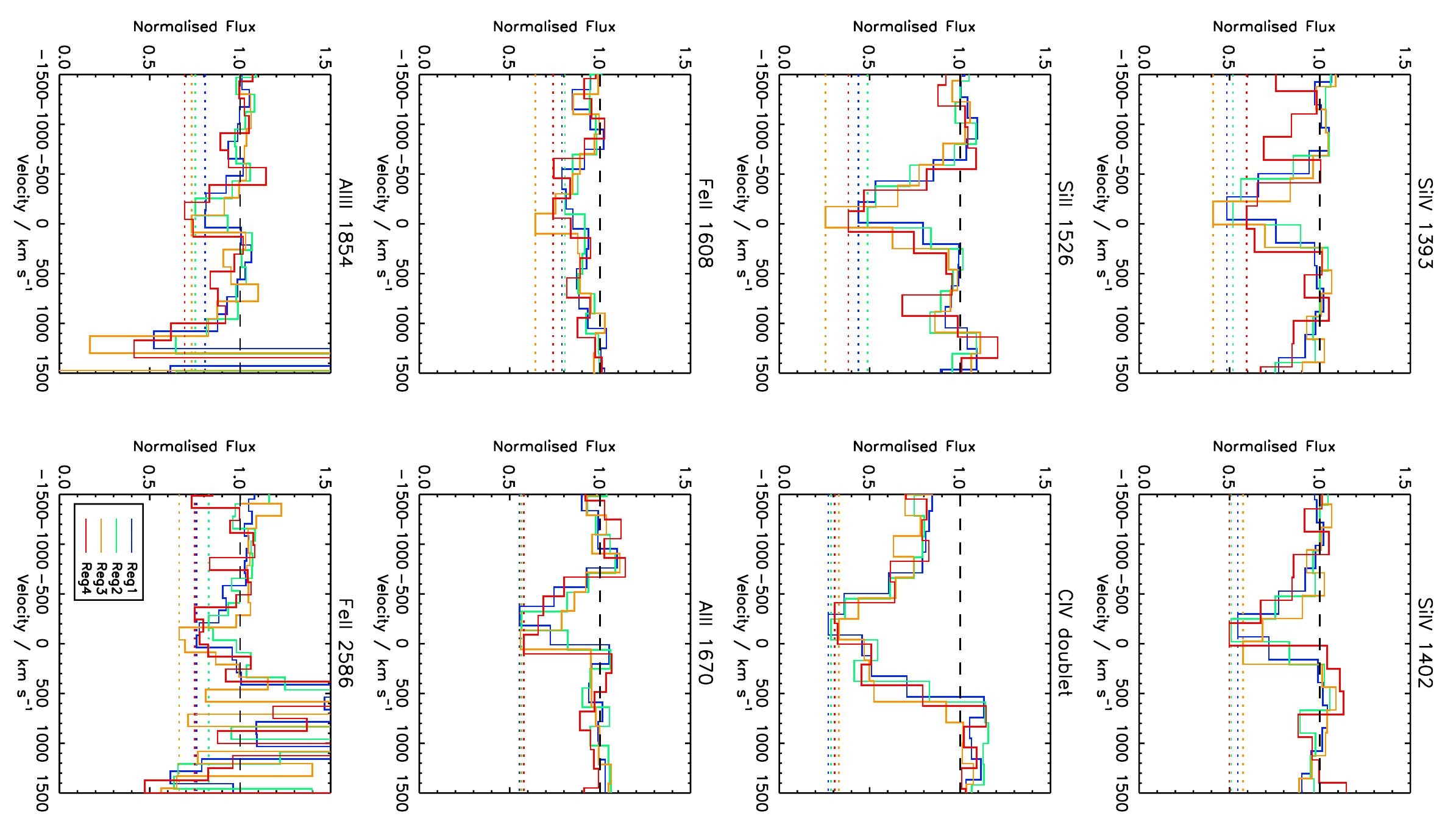}
\caption{Normalised spectra showing absorption line profiles of \sIii\,$\lambda1526$, \civ\,$\lambda\lambda1548, 1550$, \sIiv\,$\lambda\lambda 1393, 1402$, \alii\,$\lambda1670$, \aliii\,$\lambda1854$, and \feii\,$\lambda1608$ \feii\,$\lambda2586$, for the regional spectra. Black dashed line shows the continuum level, coloured dotted lines represent the profile minimum. Velocities are relative to the systemic velocity of its respective region (i.e. derived from the regional emission line redshift given in Table~\ref{tab:fluxes}).}~\label{fig:profiles}
\end{figure*}

\begin{figure*}
\includegraphics[angle=90,scale=0.7]{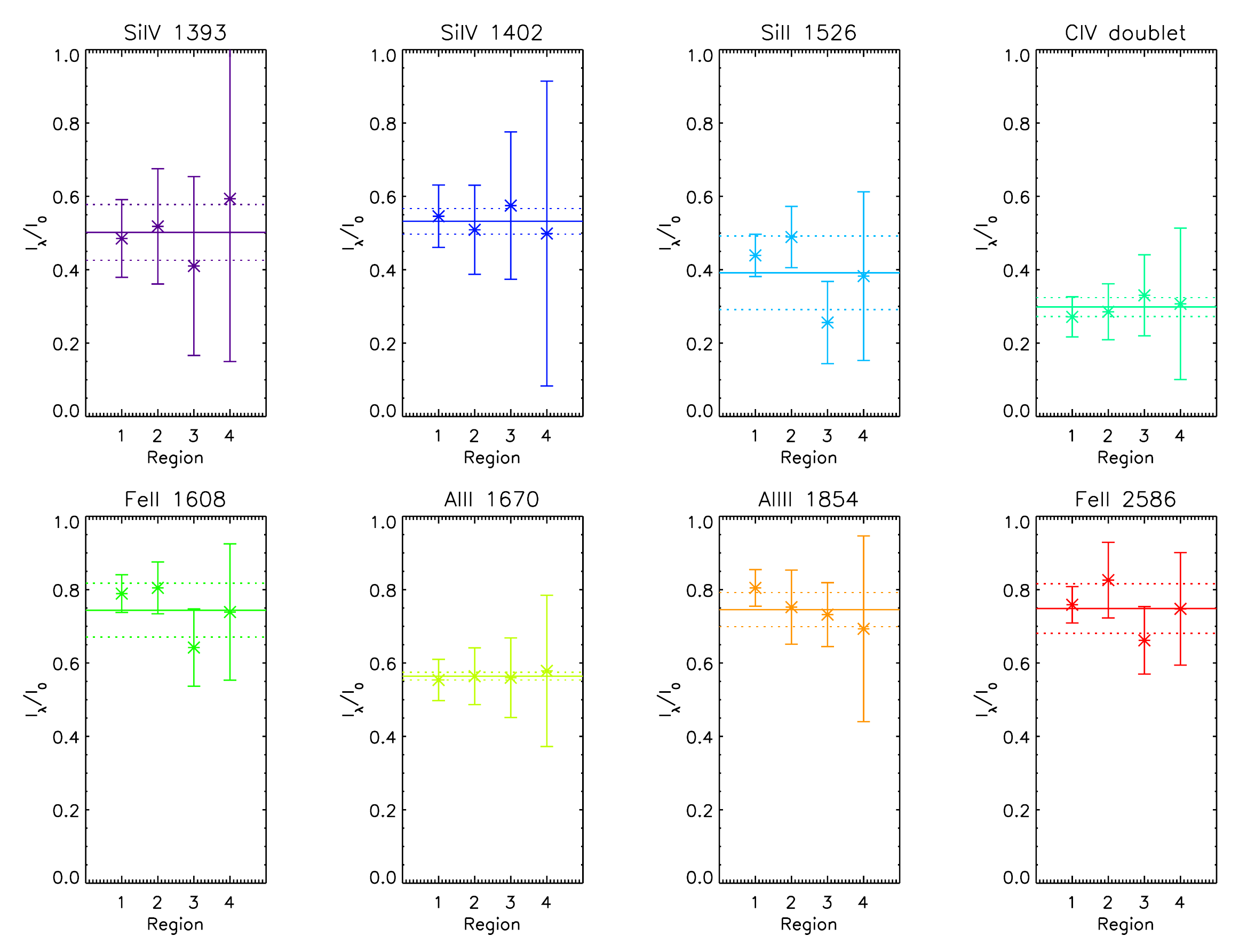}
\caption{Residual intensities for both interstellar and wind lines as a function of position across the Cosmic Horseshoe (i.e. measured from spectra extracted over the four separate regions, as shown in Figure~\ref{fig:apertures}).  Solid lines represents the mean, dashed line represents the standard error on the mean. }\label{fig:CFrac_all}
\end{figure*}

\section{Results}~\label{sec:results}
Figure ~\ref{fig:spectrum} shows the total, flux-calibrated MUSE spectrum of the Cosmic Horseshoe, obtained by summing the four regions described in Section~\ref{sec:model}. The spectrum shows all the typical interstellar absorption and nebular emission lines commonly seen in star-forming galaxies at $z = $2--3 \citep{Shapley:2003}. In this section, we describe the main characteristics of the emission and absorption lines within the MUSE wavelength range for the spectra extracted from the four source regions described in Section~\ref{sec:model}.

\subsection{Absorption line properties}\label{sec:abs}
\begin{table*}
\begin{center}\caption{Absorption line properties from regional spectra.}
\begin{tabular}{cccccc}
\hline
\input{tables/Abs_properties.tex}
\end{tabular}\label{tab:abs}
\end{center}
\begin{description}
\item \textsc{Notes:}
\item $^{\rm a}$ Rest vacuum wavelengths from \citet{Morton:2003}.
\item $^{\rm b}$ Centroid observed-frame air wavelength.
\item $^{\rm c}$ Velocity relative to the respective systemic of that spectrum, $z_{\rm em}$ , as given in Table~\ref{tab:fluxes}.
\item $^{\rm d}$ Velocity range for equivalent width measurements relative to $z_{\rm em}$, as given in Table~\ref{tab:fluxes}.
\item $^{\rm e}$ Rest-frame equivalent width and $1\sigma$ error.
\item $^{\rm f}$ Refers to \civ~$\lambda$1549 doublet.
\end{description}
\end{table*}%

\begin{figure}
\includegraphics[angle=90,scale=0.38]{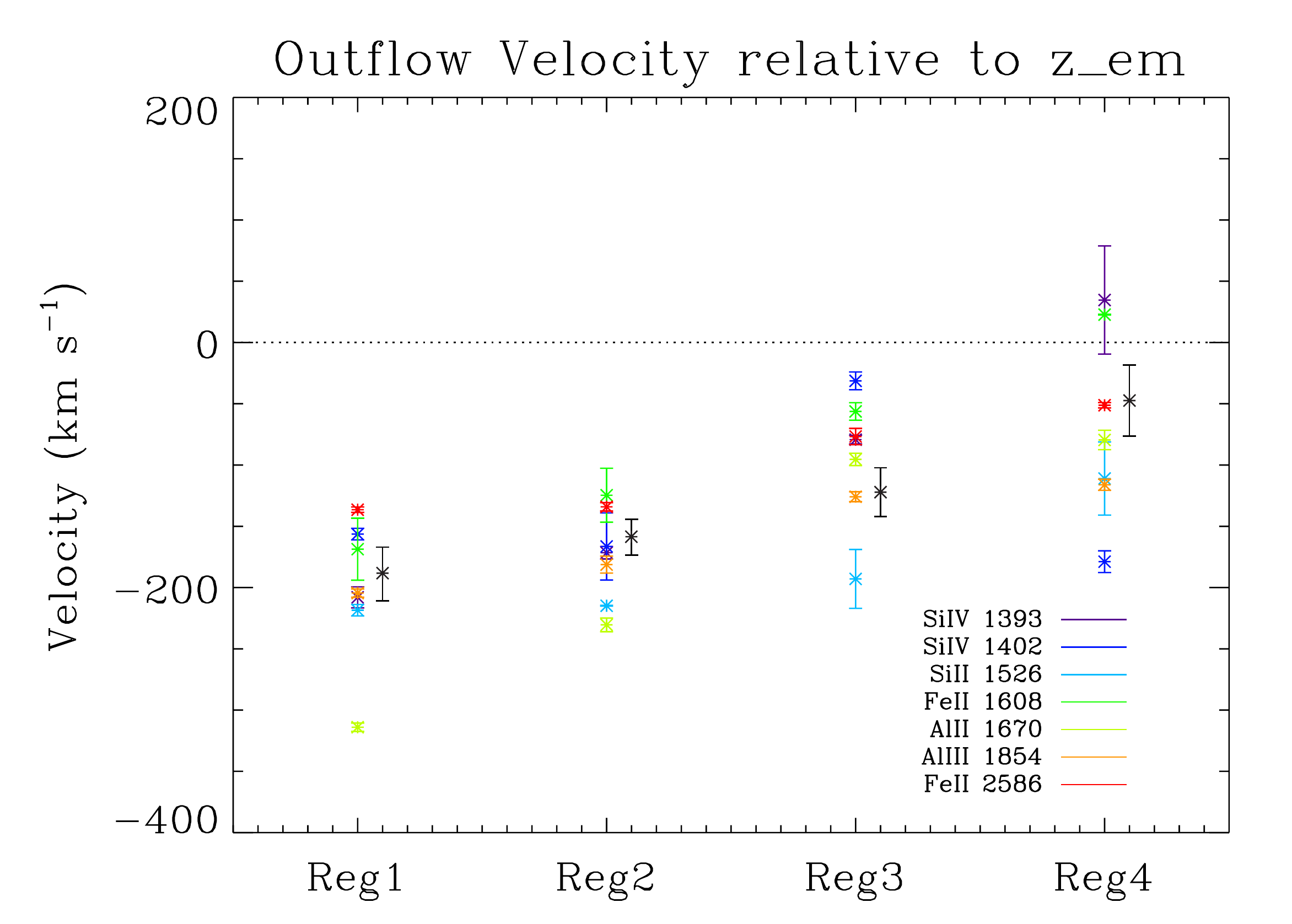}
\caption{Outflow velocities of the interstellar lines as a function of position across the Cosmic Horseshoe. The outflow velocity is defined as the velocity shift between the absorption line centroid and the systemic velocity (i.e. average emission line velocity). Black solid line represents the error-weighted mean velocity from all lines.}~\label{fig:absprop}
\end{figure}

All of the main photospheric (\sv\,$\lambda1501$), interstellar (\sIii\,$\lambda1526$, \feii\,$\lambda1608$ \feii\,$\lambda2586$, \alii\,$\lambda1670$, \aliii\,$\lambda1854$), and interstellar+wind blended lines (\civ\,$\lambda\lambda 1548, 1550$, \sIiv\,$\lambda\lambda 1393, 1402$), typically observed within UV spectra of star-forming galaxies are seen within the MUSE spectrum of the Cosmic Horseshoe, as labelled in Figure~\ref{fig:spectrum}. The properties of such lines can offer a plethora of information about the physical (e.g. outflow kinematics, covering fractions) and chemical (e.g. density, temperature, metallicity composition) of the gas and stars within the system.  Here we refrain from entering into a discussion on the existence and properties of these lines in the Cosmic Horseshoe itself and in relation to other systems, since this is proficiently covered in Q09.  Instead we exploit the spatially resolved nature of our data and concentrate on determining whether the properties of these absorption lines change as a function of position throughout the Cosmic Horseshoe. 

In Figure~\ref{fig:profiles} we show spectra for the individual regions 1-4 in the source described in Section 3, zoomed-in to show the line profiles of \sIiv, \sIii, \civ, \feii, \alii, and \aliii. Normalized spectra were created using a fixed-set of continuum nodes for each spectrum, carefully placed in feature-free wavelength regions \citep[e.g.,][]{Rix:2004}. Upon initial inspection, the line profiles between the regions appear to be very similar. We explore this further by extracting several different properties, such as the optical depth, line velocity, rest-frame equivalent widths, $W_0$, and velocity range over which the absorption takes place - all listed in Table~\ref{tab:abs}. We discuss our findings in the following subsections.
\begin{figure}
\includegraphics[angle=90,scale=0.4]{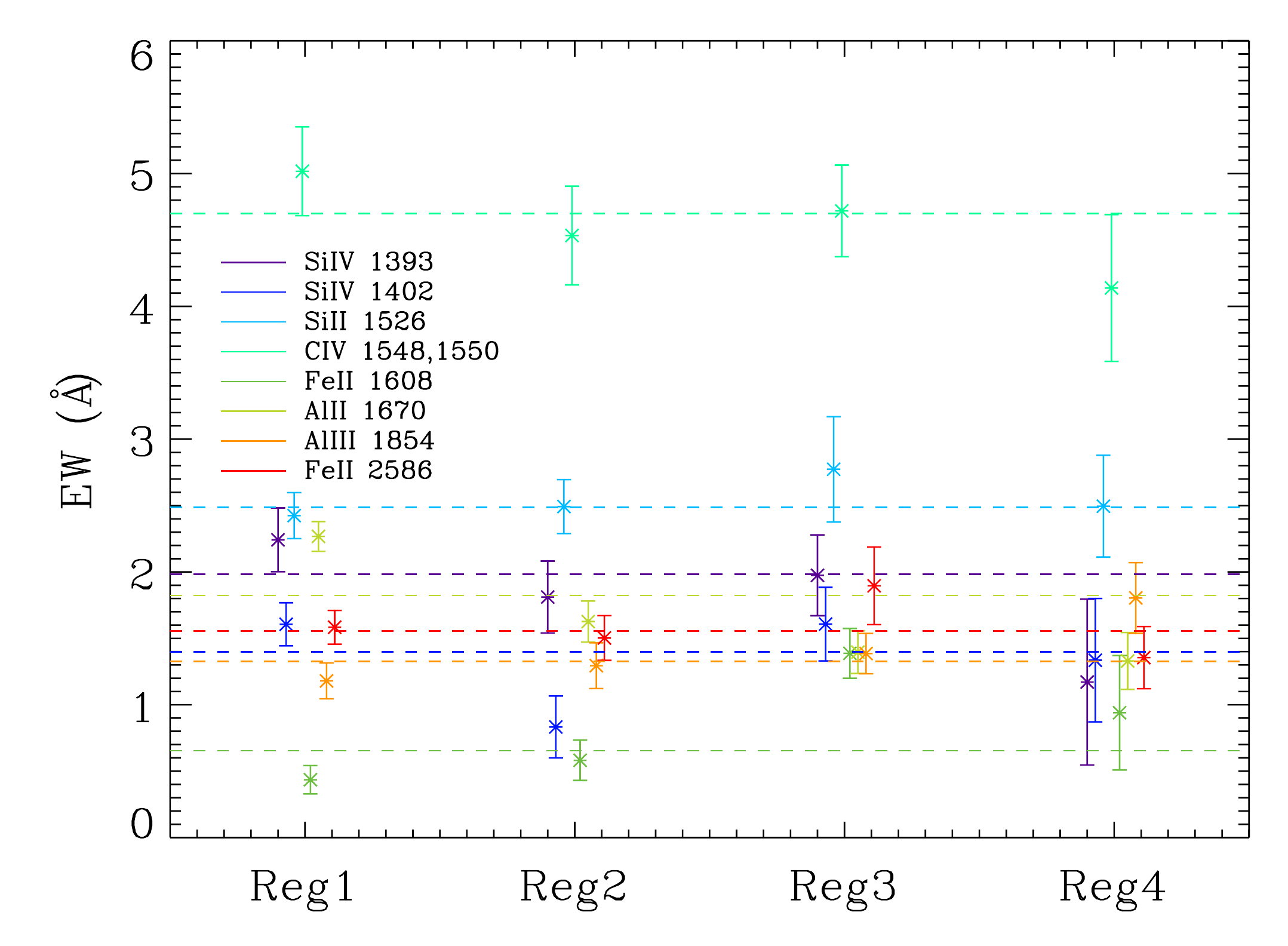}
\caption{Equivalent widths for the main absorption lines seen within the MUSE spectrum of the Cosmic Horseshoe for each of the regional spectra. Dashed horizontal lines represent the weighted mean equivalent width for that line.}~\label{fig:EW}
\end{figure}

\subsubsection{Optical Depth}
The shape and residual line intensity of an absorption line can be important measures of two quantities - the column density of the absorbing gas for that particular ion and, in the case of saturated lines, the fraction of UV flux covered by absorbing gas. As such, a variation in line depth for the same ion throughout a system would reveal a change in the amount or spatial location of absorbing gas. Unfortunately, however, we are unable to distinguish between these two cases with our MUSE data due to the relatively low spectral resolution of the data, which causes flux to be deposited from the core and into the wings of the line profile. This is demonstrated in Figure~\ref{fig:profiles}, where even the strongest absorption lines (such as the \civ~$\lambda 1549$ doublet and \sIii~$\lambda 1526$) which are often saturated in spectra of star-forming galaxies at $z = 2$--3
show a minimum apparent transmission of $I_\lambda/I_0\sim$0.3--0.4. 

What we \textit{can} assess is whether there is a change in overall optical depth throughout the four regions. In Figure~\ref{fig:CFrac_all} we show the minimum flux of each absorption profile and the uncertainty on that flux (Fig.~\ref{fig:profiles}, coloured dotted lines) for each line, for each of the four source regions. While the minimum line intensity may change for some lines (e.g. \sIii~$\lambda$1526) across different regions, we cannot say that the optical depth is consistently higher or lower in a particular region of the galaxy, within the uncertainties.

\subsubsection{Gas Kinematics}
The speed of outflowing gas, whether it be due to a galactic or stellar wind, can be determined via the minimum (i.e. blueshifted) velocity of the absorption line profile originating in that outflow.  More specifically, it is the ISM component of the line profile that directly traces the outflowing gas due to galactic winds. By measuring the velocity profiles of ISM lines at different regions throughout the Cosmic Horseshoe, we gain information on both the velocity and amount of outflowing gas as a function of position.  In order to assess the velocity of the outflowing gas, we first measure the redshift of the absorption profile, derived from the line centroid and the vacuum wavelength of the line. We then measure the minimum and maximum line velocity (i.e. the velocity at which the respective blue and red wing of the line profile returns to the continuum level). 
All velocities are defined in relation to $z_{sys}$, the systemic (i.e. emission line) redshift measured for that region (Table~\ref{tab:fluxes}).  The outflow velocity, minimum velocity and maximum velocity of each line are given in Table~\ref{tab:abs}, along with their respective uncertainties. Uncertainties in velocity measurements are derived on a `by-eye' basis by estimating the upper and lower velocity boundaries of each wing with respect to the continuum level. In the case of lines that are blends of interstellar absorption and P-Cygni profiles from the winds of luminous OB stars, namely \sIiv\,$\lambda\lambda 1393, 1402$ and \civ\,$\lambda\lambda 1548, 1550$, velocity measurements refer only to the ISM component of the line. It should be noted, however, that we do not include \civ\, in our velocity analysis, due to the large amount of blending between the components (such blending also prevents us from assessing any regional variation in the P-Cygni profile itself).

The minimum line velocity of all lines, or outflow speed of the gas, decrease from Region 1 to 4 from $\sim-800$--$\sim-400$\,\kms\ to $\sim-500$--$\sim-200$\,\kms. However, there does not appear to be any significant change in the maximum velocity of the lines between the different regions throughout the galaxy, with all ISM lines showing a maximum velocity of $\sim$50--300~\kms\ for all regions. The variation in outflow velocity can also be seen in Fig.~\ref{fig:absprop}, where we assess the velocity of the absorption lines (i.e. the velocity of the profile minimum) relative to the systemic velocity (i.e. the emission line velocity for the respective region). This is essentially a measure of the relative velocity between the outflowing ISM and the ionised gas. Again Region 1 shows the largest outflow velocity, $\sim-200$\,\kms, and Region 4 the lowest ($\sim-50$\,\kms).  Judging from the mean velocity offset for each region (black solid line) and the standard error on the mean, the change in outflow velocity between the four source regions of the galaxy is significant with respect to the uncertainties.

\begin{figure*}
\includegraphics[angle=0,scale=0.4]{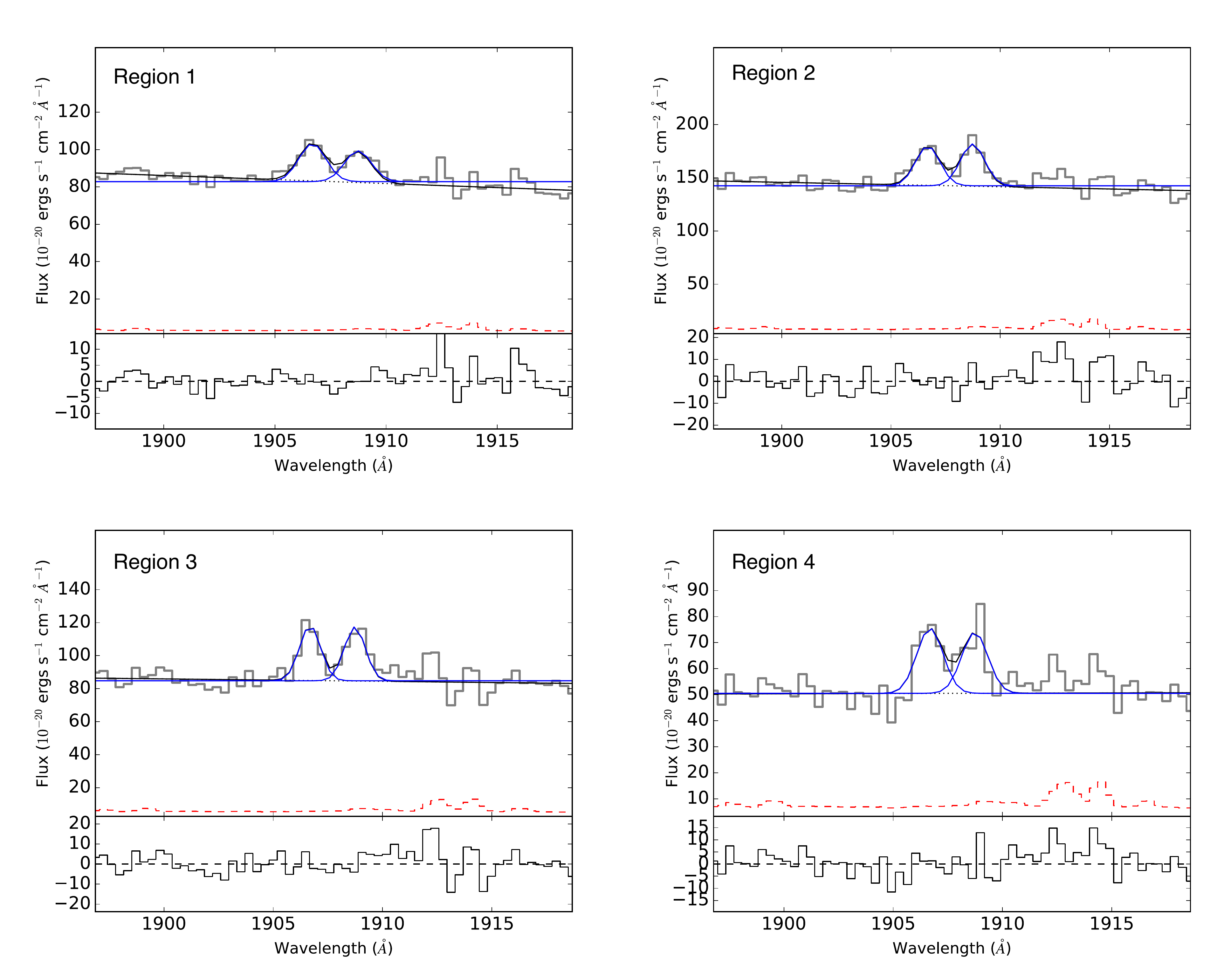}
\caption{The \fciii~$\lambda$1906 \sfciii~$\lambda$1908 emission line doublet observed within the MUSE wavelength range for the regional spectra. Each panel shows the rest-frame wavelength in \AA\ and the flux in units of 10$^{-20}$ \flux. In each plot, the grey histogram is the observed spectrum and the red dashed line is the 1$\sigma$ error spectrum. We overlay the best-fitting profiles for the emission (blue solid line) and their sum (black solid line). Relevant parameters of the model fit and line fluxes are given in Table~\ref{tab:fluxes}. Underneath each emission line we show the difference between the data and the best-fitting model.}\label{fig:emlines}
\end{figure*}

\subsubsection{Equivalent Width}
We plot the equivalent width ($W_0$) as a function of position for all absorption lines in Fig.~\ref{fig:EW}.  Equivalent widths were measured for the ISM component of the line only and in the case of \civ$\lambda\lambda 1548, 1550$, $W_0$ refers to the blended doublet. It can be seen that for \sIii, \sIiv~$\lambda$1402, \civ, \aliii~$\lambda$1854 and \feii~$\lambda$2586, $W_0$ remains constant across the galaxy within the uncertainties.  Small variations outside the weighted mean value can be seen for  \sIiv~$\lambda$1393, \alii~$\lambda1670$ and \feii~$\lambda$1608. However, for \sIiv\ and \feii, the same variation is not seen in both lines of the same species and is therefore not deemed to be definitive.

\subsection{Emission-line properties}\label{sec:emlines}
In Figure~\ref{fig:emlines} we show the \fciii~$\lambda1906$ and \sfciii\,$\lambda1908$ emission line doublet detected within the MUSE rest-frame UV spectrum, for the spectra extracted over the four main source regions. In line with previous observations (Q09), \sfoiii\,$\lambda\lambda 1606, 1666$, or \heii~$\lambda$1640, were not detected above the 3$\sigma$ level in the `total' spectrum (Fig.~\ref{fig:spectrum}) or regional spectra. 

The \fciii,\sfciii\, doublet was fitted simultaneously using two Gaussian profiles at a fixed separation in rest wavelength. The Gaussian model included two parameters (the redshift $z$ and velocity dispersion $\sigma$) that were determined by a Markov chain Monte Carlo (MCMC) method, while the fluxes in each emission component were treated as linear parameters. The continuum flux was modeled using a linear fit within a 30~\AA\ box centered on the doublet. For each MCMC step we solved the bounded linear problem that determines the (non-zero) amplitude of each line given the current values for the non-linear parameters; we explicitly accounted for the uncertainties from the linear inversion by attaching samples for the inference of the line amplitudes to the MCMC chain. The fluxes measured from each regional spectrum, along with redshift and line width, are given in Table~\ref{tab:fluxes}. 

\subsubsection{\sfciii\, Kinematics}
\begin{figure}
\includegraphics[angle=90,scale=0.8]{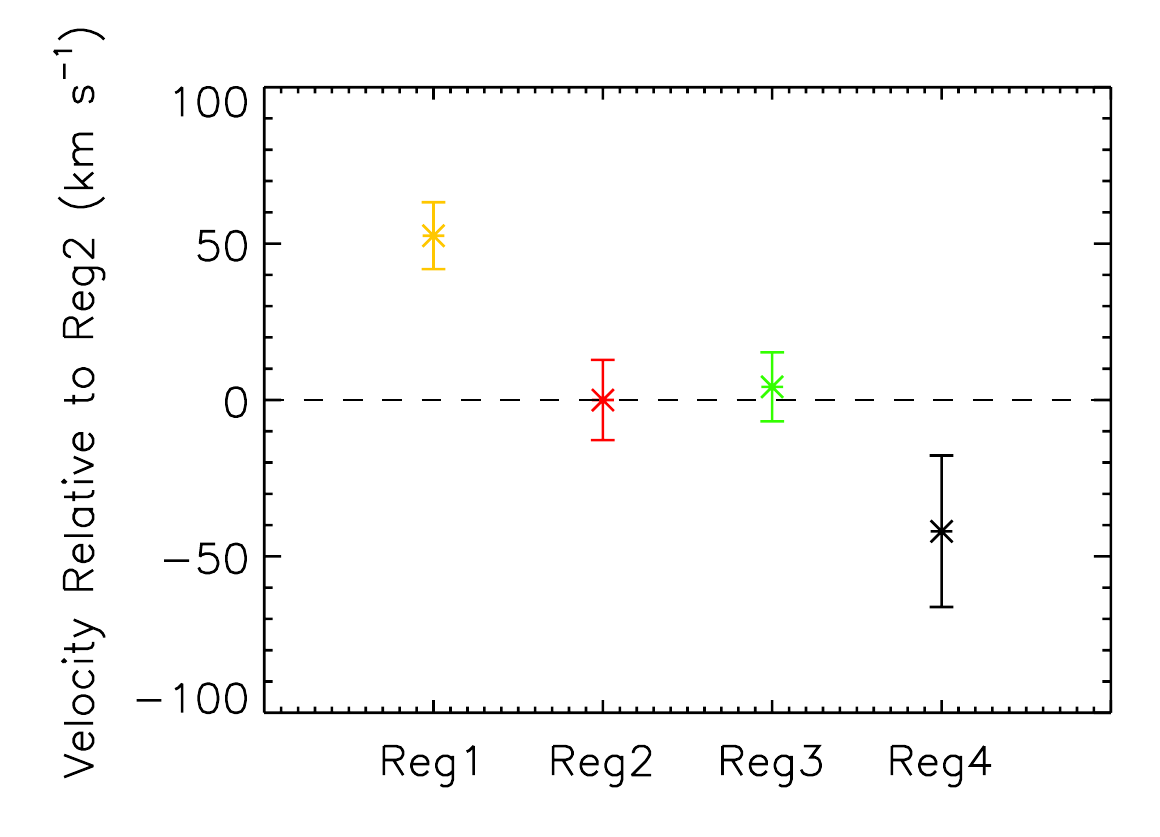}
\caption{Emission-line velocity relative to Region 2, as measured from the \fciii~$\lambda1906$ and \sfciii$\lambda1908$ emission line doublet. Points are color-coded according to the regional apertures shown in Fig~\ref{fig:apertures}.}\label{fig:redshift}
\end{figure}

The emission-line fits reveal a significant variation in redshift throughout the system. In Fig.~\ref{fig:redshift} we plot the emission-line velocity of each region relative to the main star-forming region (Region 2), showing a distinct decrease in velocity from Region~1 ($\sim50$\,\kms) out to Region~4 ($\sim-40$\,\kms).   A variation in emission line velocity was also observed by \citet{Jones:2013}, who mapped the \ha\ velocity across the system and found signs of ordered rotation and a peak-to-peak velocity shear of 148$\pm$2\,\kms.  The direction of the velocity gradient traced by the \ha\ kinematical map and the regional \sfciii\ velocities are in agreement, with both emission lines showing a positive-to-negative velocity gradient in the south (Region 1) to north (Region 4) direction, relative to Region 2.  

Velocity dispersions range from $\sim 67$ to 88~\kms\ (after correcting for an instrumental resolution of $\sim45$\,\kms\ at 1907~\AA). While the emission lines may be marginally narrower in Region 3, all four Regions are consistent with having the same velocity dispersion within the uncertainties. Conversely, the \ha\ emission map of \citet{Jones:2013} does show a significant variation in velocity dispersion between 20 and 130\,\kms\, with an average of 90$\pm$33\,\kms, across the system. As such, the lack of detected variation in $\sigma_{C\,\sc{III}]}$ is most likely due to the low S/N of the \sfciii\ doublet.

\subsubsection{\sfciii\, Equivalent Width}
The rest-frame equivalent width of the \sfciii\ doublet also shows significant variation across the four regions of the Cosmic Horseshoe (Table~\ref{tab:fluxes}). The equivalent width of the \sfciii\ doublet is thought to correlate with increasing intensity of the ionising radiation (i.e. ionisation parameter, $U$) and decreasing metallicity \citep[e.g.,][]{Rigby:2015,Maseda:2017,Nakajima:2017}. As such, strong global \sfciii\ emission (i.e. $W_0\gtrsim20$~\AA) is used to trace young, intensely star-forming galaxies at high-$z$ \citep{Stark:2014, Stark:2015}. Understanding whether such correlations hold on a \textit{spatially-resolved} basis, however, has been hindered until now by the relative faintness of the doublet. 
Here we see that while Regions 1--3 all lie within 0.6--0.8~\AA, Region 4 has a higher equivalent width by $\sim2\sigma$. This suggests that Region~4 should show signs of more intense star-formation in younger, or lower metallicity environment and, as a result, we would expect Region 4 to be bluer in colour than the other regions.  However, the $HST$ photometry (Table~\ref{tab:regions}) shows that this in fact not the case, with Region~4 actually being the reddest of the four regions and Region~2 as the bluest. Moreover, the metallicity gradient found by \citet{Jones:2013} appears to be decreasing \textit{away} from Region~4 (although this is somewhat difficult to tell without the metallicity map itself).  To fully investigate whether $W_0$(\sfciii) is indicative of intense star-formation in this system, we would need to map the ionisation parameter to see whether Region~4 shows signs of harder ionising radiation (unfortunately this property was not mapped by the \citet{Jones:2013} study because \foii\ was not within the wavelength range of their observations).  The SFR density (i.e. star-formation rate per unit area, $\Sigma_{SFR}$, as detailed in Section ~\ref{sec:SFR}) could also be used as an approximate indicator of the intensity of star-formation within each region. However, we do not see any correlation with this parameter and $W_0$(\sfciii), mostly due to the large uncertainties in $\Sigma_{SFR}$.

Overall, all regions show $W_0$ values that are relatively low for galaxies at $2<z<4$, which are on average $\sim$2~\AA\ \citep{Nakajima:2017}. However, the Cosmic Horseshoe is not considered to be `typical' of galaxies of this epoch (with respect to its relatively high stellar mass and half-solar metallicity, Q09). Indeed, \sfciii\ equivalent widths of $<2$\,\AA\ are more inline with those seen in higher metallicity, less intensely star-forming systems, as shown in the compilation by \citet{Rigby:2015}. Such systems are also found to be weak \Lya\ emitters \citep{Stark:2014}, as is true for the Cosmic Horseshoe with $W_0$(\Lya)$\sim11$ (Q09).

\subsubsection{\sfciii\, Electron Density}\label{sec:dens}
The \sfciii\ doublet is density sensitive and, as such, the ratio of the individual lines can be used to derive the electron density, \eld, throughout the Cosmic Horseshoe. Electron densities were calculated by assuming that ions are well-approximated by a 5-level atom \footnote{implemented via the IDL library \textsc{impro}: \url{https://github.com/moustakas/impro}} and using the updated atomic data presented in \citet{Berg:2015}. We assume an electron temperature, \elt, of 10,000~K, which is the typical \elt\ value for \hii\ regions\footnote{Although the Cosmic Horseshoe's $\sim$0.5~\Zsol\ metallicity may imply that a higher \elt\, should be used here, without knowledge of the ionisation conditions within the gas, we cannot be certain. Moreover, the \sfciii\ diagnostic has very little dependence on \elt\ \citep{Maseda:2017}.}.  The \eld\, values measured from each regional spectrum are listed in Table~\ref{tab:fluxes}.

The electron density appears to be constant throughout the galaxy, with densities ranging between $\log$(\eld)$=3.92$--4.36~cm$^{-3}$. The  regional densities all agree within the uncertainties and align well with the \sfciii -based estimate by Q09, of \eld\, 5000--25,000~cm$^{-3}$. These are higher than `typical' \hii\ region densities (usually around $\sim$100~cm$^{-3}$) both locally and around $z=$2--3 \citep[e.g.][]{Masters:2014, Sanders:2016, Kaasinen:2017}, where measurements are made using the density sensitive doublets in the optical regime (e.g. \foii\, and \fsii ). However, this value is quite typical of cases which utilize the \sfciii\ doublet \citep[e.g.][and references therein.]{James:2014} and the discrepancy between the two diagnostics may be due to the fact that this particular doublet traces the `medium ionization zone' rather than the `low ionisation zone', with the former inherently having a higher electron density. We discuss this discrepancy further in the following paragraphs.

\begin{table*}
\caption{Emission line properties of the \sfciii\ doublet measured from the regional spectra, and electron densities calculated from the corresponding ratios (as detailed in Section~\ref{sec:dens}). }
\begin{center}
\begin{tabular}{|cccccccc}
\hline
Spectrum & z & $\sigma^a$ (\kms) & $W_0^b$(\AA) & $F_{1906} ^c$ & $F_{1909}^c$& \sfciii\ ratio & $\log$(\eld (cm$^{-3}$))\\
\hline
\input{tables/logdensity.tex}
\hline
\end{tabular}
\end{center}
\label{tab:fluxes}
\begin{description}
\item \textsc{Notes:}
\item $^{\rm a}$ Velocity dispersion corrected for instrumental resolution.
\item $^{\rm b}$ Rest-frame equivalent width of \fciii~$\lambda1906$ + \sfciii\,$\lambda1908$ and $1\sigma$ error.
\item $^{\rm c}$ In units of  $10^{-20}$\,erg\,s$^{-1}$\,cm$^{-2}$\ \AA$^{-1}$.
\end{description}
\end{table*}%

At first sight, the electron densities measured here (\avgdens),  may appear surprisingly high compared to typical \hii\ region electron densities ($\sim100$~cm$^{-3}$). However, such densities are typical of galaxies at this epoch when measuring the electron density via the \sfciii\ doublet. It is interesting to note that in the handful of cases where both the rest-frame optical (e.g. \foii, \fsii ) and rest-frame UV (\sfciii) electron density diagnostics are available, the \eld\, values derived using the latter are almost consistently higher by two orders of magnitude. This is true both at $z=$1--3 \citep{James:2014, Christensen:2012, Bayliss:2014} and in the local Universe \citep{Berg:2016} - although this constitutes only 6 cases in total, including the Cosmic Horseshoe. It should also be noted that \citet{Patricio:2016} found similar values of \eld\ derived from the \sfciii\ and \foii\  density diagnostics for a star-forming galaxy at $z=3.5$, with both showing $\sim100$~cm$^{-3}$, i.e. typical of more local \hii\ regions. However, the authors \textit{do} note a large discrepancy in \eld\ values between these diagnostics and two other rest-frame UV diagnostics, \sfniv~$\lambda\lambda$1483, 1486 and \sfsIiii~$\lambda\lambda$1883, 1892, and suggest that such discrepancies may be due to local variations in density and temperature within a galaxy, and/or the fact that N$^{3+}$ is probing higher ionization zones.

Understanding the true root of these differences is of particular importance to high-$z$ studies, when we begin to rely heavily on rest-frame UV diagnostics. We propose here that the main cause for the offset between the \sfciii\ and \foii\ \eld\ values is simply that the \sfciii\ doublet is tracing regions of higher density than the optical doublets. The critical density of each line of density-sensitive doublets dictates the range between the low- and high-density limits that it is sensitive to. For \foii, the critical densities imply a transition between $\sim30$~cm$^{-3}$ and $\sim1.6\times10^4$~cm$^{-3}$ whereas for \sfciii\, the low density limit lies at $10^2$~cm$^{-3}$ and the high density limit lies at $10^7$~cm$^{-3}$ \citep{Osterbrock}. 
Since the \sfciii\ doublet transitions have higher critical densities than those in \foii\ or \fsii, it can be used as a density diagnostic in denser environments than \foii\ or \fsii.
This information supports the notion that \sfciii\ traces densities in the medium-ionisation, rather than low-ionisation zone. As such, we would expect the peak in \sfciii\ emission (especially in the \sfciii~$\lambda$1908 line, since \eld\ increases with the 1906\,\AA/1908\,\AA\ line ratio) to be closer to the ionising source than the optical doublet emission. One way to test this hypothesis would be to perform high spatial resolution mapping of the doublet. However, this is currently rather difficult due to the fact that IFU observations of this doublet are only available for objects at $z\gtrsim$0.8, where a spatial resolution of 0.5\arcsec\ corresponds to $\sim$4~kpc (in un-lensed objects). Of course, either nearby or highly magnified lensed systems would provide much finer spatial scales and an excellent way to test this hypothesis, if both the optical and UV diagnostic ratios could be mapped within the same system.

\begin{figure*}
\includegraphics[angle=90,scale=0.7]{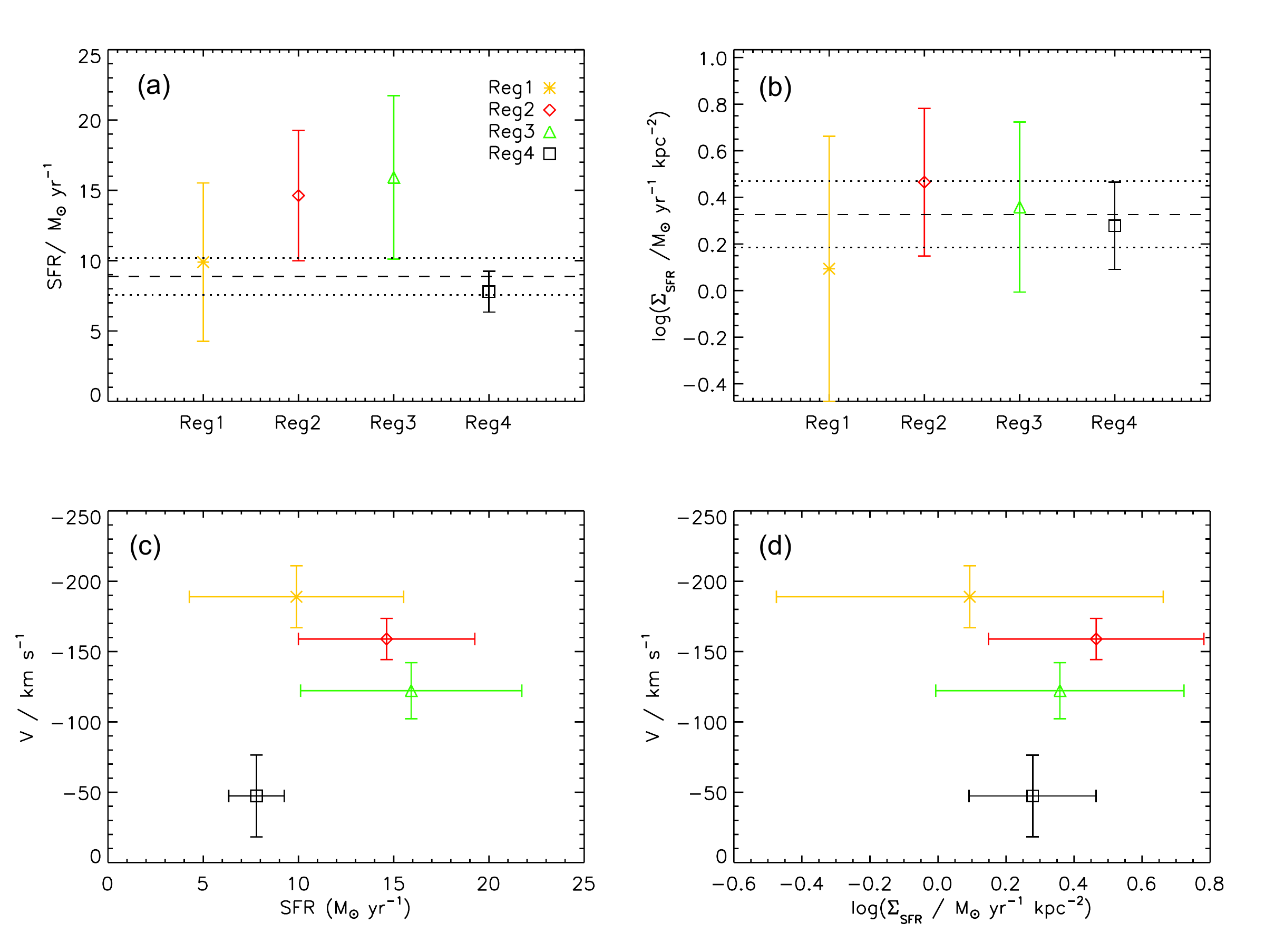}
\caption{\textit{Top panels:} Star-formation rates (\textit{a}) and star-formation rate densities (\textit{b}) measured from the regional spectra across the Cosmic Horseshoe as a function of position (points are color-coded according to the regional apertures shown in Fig~\ref{fig:apertures}). In panel-(a), dashed line represents the weighted mean value for Regions 1--4, $<\rm{SFR}>=8.9\pm1.3$~\Msol\, yr$^{-1}$, dotted lines represent the 1$\sigma$ variance. In panel-(b) dashed line represents the weighted mean value for Regions 1--4, $<\Sigma_{SFR}>=2.1\pm0.3$~\Msol\, yr$^{-1}$\,kpc$^{-2}$, dotted lines represent the 1$\sigma$ variance.  \textit{Bottom panels}: outflow velocity of the ISM relative to $z_{sys}$ as a function of star-formation rate (\textit{c}) and star-formation rate density (\textit{d}).}\label{fig:SFR} 
\end{figure*}

\begin{table*}
\begin{center}\caption{Reddening and UV SFRs calculated from regional spectra.}
\begin{tabular}{ccccccc}
\hline
Spectrum & $E(B-V)$ & $F_{1500}$   $\times10^{-20}$ & $I_{\nu,1500}$ $\times10^{-30}$& $L_{\nu,1500}$  $\times10^{29}$  & SFR(UV) & $\log(\Sigma_{SFR})$ \\
 &  &
 (ergs s$^{-1}$ cm$^{-2}$ \AA$^{-1}$ )& 
  (ergs s$^{-1}$ cm$^{-2}$ Hz $^{-1}$) & 
(ergs s$^{-1}$ Hz$^{-1}$) &  
 (\Msol\,yr$^{-1}$)  & 
 (\Msol\,yr$^{-1}$ kpc$^{-2}$ )\\
  \hline
\input{tables/SFR_tab.tex}
\hline
\end{tabular}\label{tab:SFR}
\end{center}
\end{table*}%
\subsection{Reddening and Star-Formation Rate}~\label{sec:SFR}

Despite not having access to the Balmer line series, we are still able to determine the star-formation rate (SFR) via the luminosity of the UV continuum produced by OB stars. However, in order to measure the true luminosity we first need to correct for reddening. We calculate the attenuation by first comparing the slope, $\beta$, of the UV continuum between 1400 and 1800~\AA\ in F$_\lambda$, with that of the theoretical continuum slope ($\beta_0$)  calculated from a \textsc{Starburst99} \citep{Leitherer:1999} model to give $\Delta\beta$ for each spectrum. The model spectrum corresponds to a Salpeter IMF, continuous SFR, mass cut-off of 1--100\,\Msol, metallicity of 0.5~\Zsol\ (Q09), at an age of 100~Myr (after which the UV slope does not change), and gives $\beta_0=-2.36$. We then convert from  $\Delta\beta$ to $E(B-V)$ using the \citep{Calzetti:2000} attenuation curve:
\begin{equation}
E(B-V)=\Delta\beta \times \log(1800/1400) \times (0.4(k_{1400} - k_{1800}))^{-1}
\end{equation}
where $k_{1400}$ and $k_{1800}$ are the attenuation curve values at 1400 and 1800~\AA, respectively. $E(B-V)$ values for each source region are given in Table~\ref{tab:SFR}, and range from 0.30 to 0.43. Uncertainties on $E(B-V)$ were found by repeating the above calculation with two small ($\sim$100~\AA) windows around 1400~\AA\, and 1800~\AA. The $E(B-V)$ values derived here suggest a relatively large amount of extinction within this system.  For comparison, \citet{Hainline:2009} derived the reddening using two methods - firstly, from SDSS photometry which yielded a modest reddening of $E(B-V)=0.15$ and secondly, from the \ha/\hg\ Balmer decrement which yielded $E(B-V)=0.45$.  It can be seen that our UV continuum measurements agree better with that derived from the Balmer decrement than the SDSS photometry, where the latter is most likely the least direct method of the three.

The flux (and its associated error) at 1500~\AA, measured by averaging over a 40~\AA\ window centered at 1500~\AA\ away from obvious emission and absorption lines, is also given in Table~\ref{tab:SFR}, along with the continuum luminosity density at 1500~\AA, after correcting for the attenuation and adopting the conventional cosmological parameters ($\Omega_M = 0.3$, $\Omega_\Lambda=0.7$, $h=0.7$).  Uncertainties in $E(B-V)$ are propagated accordingly into the de-reddened flux at 1500~\AA\, and as such become the dominant source of error in $I_{\nu}$ (and subsequent parameters derived from it).   We then convert into a SFR using:
\begin{equation}
\rm{SFR(UV)}=1.4\times10^{-28}L_\nu \times \frac{1}{1.8}
\end{equation}
where the first factor on the right-hand side of the equation is the calibration of \citet{Kennicutt:1998} appropriate to a Salpeter IMF, corrected for the more realistic turnover at low masses proposed by \citet{Chabrier:2003} (amounting to a factor of 1.8).\footnote{A correction is also typically made here for magnification although in this case our spectra have already been normalised to the de-lensed flux (Section~\ref{sec:model}).} 

SFRs for each source region are listed in Table~\ref{tab:SFR} and plotted in Figure~\ref{fig:SFR}a.  We can see that the SFRs do not vary across the different star-forming regions within the uncertainties, with each region measuring between $\sim$8 and 16~\Msol\,yr$^{-1}$, with a weighted-mean SFR of $\sim9\pm1$~\Msol\ yr$^{-1}$.  The lack of variation in SFR is somewhat unsurprising given the relatively constant surface brightness distribution around the Einstein ring (Fig.~\ref{fig:colorFig}), combined with the large uncertainties in reddening estimates. Unfortunately, \citet{Jones:2013} do not derive a SFR map from their \ha\ emission line maps that we could use for comparison here. The total SFR measured from summing all regional SFRs is $\sim48\pm9$~\Msol\,yr$^{-1}$. Although high compared to SFRs measured in the local Universe, this value is typical of star-forming galaxies at this redshift, i.e. during the `peak of cosmic star-formation'.  This value agrees with that of Q09, who measured $\sim$56~\Msol\,yr$^{-1}$ and within one-sigma of the 210$\pm$167~\Msol\,yr$^{-1}$ reported by \citet{Jones:2013} (although it should be noted here that the magnification factor may have been overestimated by those authors, as discussed in Section~\ref{sec:model}).

Table~\ref{tab:SFR} also lists the SFR density, $\Sigma_{SFR}$, defined as SFR divided by the area of each region (in kpc$^2$, Table~\ref{tab:regions}). SFRDs vary from $\sim$1.2 to 3.0~\Msol\,yr$^{-1}$\,kpc$^{-2}$ between the source regions and remain relatively constant across the system within uncertainties (Fig.~\ref{fig:SFR}c), with an average of $\Sigma_{SFR}=2.1\pm0.3$~\Msol\,yr$^{-1}$\,kpc$^{-2}$. By comparison, the SFRD for the entire system is also $\sim$2~\Msol\,yr$^{-1}$\,kpc$^{-2}$ (derived from a total SFR of 50~\Msol\,yr$^{-1}$ and total area of 24~kpc$^{-2}$).  As expected from Fig.~\ref{fig:apertures}, the most diffuse source region, Region 1, has the lowest SFRD, whereas the most compact region, Region 2, has the highest SFRD. It should be noted that all of the individual star-forming regions possess a SFRD that is well above the 0.1~\Msol\,yr$^{-1}$\,kpc$^{-2}$ SFRD threshold given by \citet{Heckman:2002}, above which starbursts are capable of driving `superwinds'. This threshold holds for both local starburst galaxies and high-$z$ star-forming galaxies, within which these starburst-driven superwinds are ubiquitous and can drive interstellar gas at speeds up to 1000~\kms. 

Such `superwinds' can play an important role in both the mixing of metals produced by nucleosynthesis within the interstellar medium, along with the regulation of star-formation itself. In order to assess the effect of the latter, i.e. whether SFRs affect the speed of outflowing gas within each region, in Fig.~\ref{fig:SFR}c and Fig.~\ref{fig:SFR}d we plot SFR and SFRD vs. outflow velocity (i.e. the average `regional' absorption line velocity relative to the systemic velocity, Table~\ref{tab:abs}). Several studies have shown that outflow velocity increases with SF activity \citep[e.g.][]{Heckman:2004,Martin:2005,Heckman:2015,Cicone:2016}, which is to be expected given that the mechanical energy driving the outflows is from massive stars in the form of supernovae and stellar winds \citep{Leitherer:1995}. Unfortunately, we see no such correlation here for either parameter (irrespective of their uncertainties), which is unsurprising given the relative agreement between SFR and SFRD between the regions. However, this lack of correlation with outflow velocity may also be because the dependencies do not hold on a spatially-resolved basis within a system. Such correlations can also depend strongly on the line in question, whether one uses the maximum velocity or line centroid, and of course the geometry of the system in question. As such, several other studies have also struggled to find a strong correlation between outflow velocity and SFR or SFRD, \citep[e.g.][]{Kornei:2012, Rubin:2014,Chisholm:2015}.

\section{Discussion and Conclusion}\label{sec:disc_conc}
In this study we have presented the first spatially resolved rest-frame UV view of the Cosmic Horseshoe using VLT-MUSE. The primary goal of our observations was to discern whether any UV properties exhibit spatial differences, such as the spatial extent, velocity, and magnitude of outflowing gas, and connect such variations to the physical conditions of the star-forming region from which the outflowing gas originates, such as star-formation rate and its density.

From our gravitational lens model, we decipher the Cosmic Horseshoe as being made up of four distinct source regions. As a result of this, rather than tracing properties continuously throughout a single system, we instead use spectra integrated over each of these four regions to understand whether we can detect UV-spectroscopic variations in systems of this kind. Morphologically, Region 1 has a filamentary structure, Region 2 is an elongated structure that holds two high surface brightness peaks (one of which is the highest within the system and as such we deem this the `main star-forming region'), Region 3 consists of two relatively low surface brightness components, and Region 4 is a single component with the second highest surface brightness. We summarise our findings about these regions below.

\subsection{Variations in Gas Kinematics}
We can see that the system has significant kinematical structure, suggesting that the four components are in the process of merging. The velocities of Regions 1 and 4 are offset from the main source region (Region 2) and Region 3 by $\sim\pm$50\,\kms. Such a fragmented structure is typical at $z=$2--3 \citep{Law:2007}. 

The interstellar lines show a variation in outflow velocity throughout the system, both with regards to the velocity of the profile minimum and the minimum velocity of the absorption profile. Region 1, despite being diffuse in nature, appears to have the strongest galactic outflow of the system, with $<v>\sim-200$\,\kms. Region 2, perhaps unsurprisingly due to its high surface brightness, also shows signs of fast outflowing gas. However, Region 4, the second brightest region, shows signs of a very weak galactic outflow ($\sim-50$\,\kms). 
While outflow velocities of $-100$ to $-300$\,\kms\ are typical of the \textit{global} outflow rates seen in galaxies at this redshift \citep[e.g.][]{Shapley:2003,Steidel:2010}, here we instead appear to be detecting \textit{localised} outflows from each of the merging components. To our knowledge, this is only the second documented case of localised outflows - the first being in the lensed galaxy, RCSGA 032727-132609 \citep{Bordoloi:2016}, where outflows with velocities of $-170$ to $-250$\,\kms\ were detected along four distinct lines of sight separated by up to 6~kpc. Somewhat smaller separations are traced here, with a larger range in outflow velocity.  Interestingly, \citet{Bordoloi:2016} find a (somewhat weak) correlation between outflow velocity and star-formation rate density and as such they deem the outflows to be `locally sourced', i.e. where outflow properties are controlled by the star-forming clump from which it originates. However this does not appear to be the case in the Cosmic Horseshoe, where we instead detect the highest outflow rate from the most diffuse star-forming region (Region 1). This suggests that star-formation may not be the only factor influencing the outflowing gas within this system. Of course, connecting small scale properties (e.g. star-formation) with large-scale gas outflows may also be an issue due to geometric effects, which we discuss in Section~\ref{sec:lessons}.

\subsection{Variations in Gas Properties}
There were no signs of variation in the strength of the interstellar absorption lines throughout the system. This could be due to the fact that the amount of absorbing gas is relatively constant throughout, as may be the case if the absorbing gas were located well in the foreground of all four clumps. An alternative possibility is that our spectra do not have the signal-to-noise ratio required to detect such variations. We were also unable to detect any changes in the optical depth of the line profiles, within the uncertainties. Moreover, due to the low spectral resolution of our data, even if such variations did exist in the saturated line profiles, we would not be able to decipher whether the cause is due to covering fraction or column density. This is particularly frustrating in the case of the Cosmic Horseshoe because previous studies of the Cosmic Horseshoe found a puzzling discrepancy between a less than unity covering factor of the absorbing gas (Q09) and a very low escape fraction of Lyman continuum photons \citep{Vasei:2016}. Spatially resolving the interstellar absorption had the potential of addressing whether a non-uniform coverage of the UV flux across the system contributes to this discrepancy. 

Our observations revealed a somewhat surprisingly constant SFR across the system, with an average SFR of $\sim13$\Msol\,yr$^{-1}$. Although we do see a slight change in SFR according to surface-brightness (Fig.~\ref{fig:apertures}), the large uncertainties in reddening prevent us from discerning a significant variation. Consequently, no significant variation in SFRD is seen either and as such we are unable to deduce any connection between outflow velocity and star-formation, as discussed above.  Both the localised SFRs and total SFR lie within the typical range for galaxies of this epoch, which can be anywhere in between 1 and 200~\Msol\,yr$^{-1}$ \citep[e.g.,][]{Masters:2014,Steidel:2014}.

For the first time, we spatially resolve the \sfciii\ equivalent widths across a galaxy. Interestingly, the strength of the \sfciii\ emission line doublet was found to vary across the four star-forming regions, with Region 4 showing a higher $W_0$(\sfciii) by $\sim2\sigma$ compared to the other regions within the system. This suggests that Region~4 may have a harder ionising radiation field than the other three regions. However, a direct measurement of the ionisation parameter, $U$, would be needed to confirm this (e.g. via spatially resolved optical emission lines such as \foiii\, and \foii).

The electron density in each of the regions (as traced by the ratio of the \sfciii\ doublet lines) was found to be the same within the uncertainties, most likely due to the relatively low S/N in emission line flux. The velocity dispersion of the gas within each region (as traced by the \sfciii\ emission lines) does not show significant variation, which may be due to the fact that any expected variations would be over a limited dynamic range that is undetectable here due to the resolution of our observations. Indeed spatial variations in $\sigma$ were detected by \citet{Jones:2013} via the \ha\ emission line, although we cannot assume that the \ha\ and \sfciii\ emission would be co-spatial within the gas. On the other hand, this is most likely the case, given the similarities in the \ha\ and continuum source plane maps.

\subsection{Lessons for future spatially resolved UV studies}\label{sec:lessons}

Finally, we address the question: what have we learnt overall about spatially resolving the UV continuum of galaxies at this epoch? 
To summarise, in this specific system, we have detected variations in the kinematics of the gas - both with regards to the intrinsic velocity of the star-forming clumps and the gas outflowing from them - and the hardness of the ionising radiation (as traced by $W_0$(\sfciii)). We did not detect variations in the amount of absorbing gas, electron density, or the star-formation rates, between each of the regions. 
 
Upon reflection, our experiment with spatially resolving the UV continuum suggests that large scale variations in gas kinematics (e.g. outflows) can be traced, while inhomogeneity in the physical conditions of the gas may be harder to detect. As such, connecting gas properties across different scales, e.g. large scale gas outflows with the properties of small scale star-forming regions, may prove difficult.  This may be due to several reasons (i) some of the physical properties studied here (e.g. electron density, gas column density) may not exhibit large variations across a system; (ii) if subtle variations do exist in these properties, high S/N data are required to detect them; and (iii) geometric effects.

With respect to the final point, understanding the spatial connection between outflows and star-forming regions, i.e. the degree to which outflows are `locally sourced', may be intrinsically difficult simply because of the large dependence on the inclination of the system. For example, if we were to observe a galaxy face-on, then the detection of varying amounts and speeds of outflowing gas and/or gas coverage, and its connection to localised SF properties, would be far easier to detect than in a system where inclination angles can `mix' the amount of absorption along the line-of-sight \citep[e.g.,][]{Bordoloi:2014}. 

The present study has shown that detecting significant variations in absorption or emission line properties within the rest-frame UV requires high signal-to-noise data, which can be achieved via the target observed or instrument used. The systematic success of such observations does depend largely on the chosen target and due to the compact and faint nature of galaxies at this epoch, we are largely constrained to using gravitationally lensed systems. Moreover, due to complications in spectral extraction within the image plane, we are also constrained to using systems that are physically distinguishable in the image plane even after accounting for the PSF. As such, in order to maximise the S/N in the UV continuum, gravitationally lensed targets for studies of this kind should (i) have a well-constrained model from high spatial resolution data, that can trace into clearly separated components in the image plane; (ii) have a high magnification; and (iii) be targets within which we would expect to see variation (i.e. systems showing sign of recent star-formation and disturbed morphology). 

In relation to the original goals of this study, the data presented here have opened up several questions concerning the spatial dependence of outflows on localised properties of the UV emitting region. While we know that a relation exists between the outflow strength and the amount of star-formation within a system as a whole, we remain unaware as to whether such relationships exist locally. For example, here we have a case where the most diffuse star-forming region shows signs of the strongest outflow, suggesting that the connection between the density of star-formation and outflowing gas is not clear cut. This is partly explained by the geometry of this particular system - if the gas is located far away from the UV emitting region, any kinematical signatures from the regions themselves would be lost. As such, perhaps we will only see kinematical signatures from the star-forming regions if the absorbing gas is located in close proximity, as is the case for RCSGA 032727-132609? Since kinematical signatures within the outflowing gas do still exist within the outflowing gas here, what \textit{other} mechanisms are affecting the magnitude and extent of the outflowing gas on a localised basis?  Of course, we cannot answer such questions from only two targets and more studies of this kind are needed. In particular, studies that combine spatially resolved rest-frame UV and optical data would be ideal in order to pair outflow velocities with robust SFR measurements from the Balmer emission lines.

On a final note, studies of this kind will be greatly assisted in the future via the increased spatial resolutions of both current and future observatories, which will help minimise the amount of source blending in the source plane.  With the recent commissioning of the adaptive-optics channel on MUSE, we should be able to achieve spatial resolutions of 0.4\arcsec. While with space-based IFU capabilities, such as the IFU mode JWST/NIRSpec, point spread functions of the order 0.1\arcsec\ will be achievable.  Furthermore, with its 0.6 to 5 $\mu$m wavelength range, JWST will cover the rest-frame UV of galaxies at $z>$4--5, where observations of faint, spatially extended stellar continuum may finally be within reach.  

\section*{Acknowledgments}
We are grateful to the European Southern Observatory time assignment committee who awarded time to this programme and to the staff astronomers at Paranal who conducted the observations. We also extend our thanks to the reviewer, Johan Richard, for his insightful and valuable comments on our manuscript. The authors are sincerely grateful to Danielle Berg for discussions concerning electron density diagnostics, along with Jane Rigby and Jason Tumlinson for discussions on spatially resolving outflows. BLJ thanks support from the European Space Agency (ESA) and SC acknowledges financial support from the Science \&\ Technology Facilities Council (STFC). The research leading to these results has received funding from the European Research Council under the European Union's Seventh Framework Programme (FP/2007-2013)/ERC Grant Agreement no. 308024. 
\bibliographystyle{mn2e}
\bibliography{references}
\clearpage

\bsp

\label{lastpage}

\end{document}

%% file: tables/Abs_properties.tex
Line ID & $\lambda_{lab}^a$ (\AA) & $\lambda_c^b$ (\AA) & $v^c$ (\kms ) & $\Delta v^d$ (\kms) &  $W_0^e$ (\AA) \\ \hline
\multicolumn{6}{c}{Reg1}\\ \hline
\sIiv $\lambda$1393 & 
 1393.7603 & 
      4709.03 & 
        -208$\pm$           8 & 
        -685$\pm$          40 to +         256$\pm$          73 & 
  2.242$\pm$  0.240 \\
\sIiv $\lambda$1402 & 
 1402.7729 & 
      4740.30 & 
        -156$\pm$           4 & 
        -491$\pm$          13 to +         230$\pm$          53 & 
  1.606$\pm$  0.162 \\
\sIii $\lambda$1526 & 
 1526.7070 & 
      5158.03 & 
        -218$\pm$           4 & 
        -664$\pm$          72 to +         227$\pm$          46 & 
  2.424$\pm$  0.173 \\
\civ $\lambda$1548 & 
 1548.2041 & 
      5231.79 & 
--- & 
--- & 
  5.017$\pm$  0.334$^f$ \\
\civ $\lambda$1550 & 
1550.7812 & 
      5231.79 & 
--- & 
--- & 
  5.017$\pm$  0.334$^f$ \\
\alii$\lambda$1670 & 
 1608.4510 & 
      5435.11 & 
        -168$\pm$          25 & 
        -394$\pm$          40 to +         143$\pm$          67 & 
  0.436$\pm$  0.107 \\
\feii $\lambda$1608 & 
 1670.7886 & 
      5643.01 & 
        -313$\pm$           3 & 
        -796$\pm$          82 to +         121$\pm$          48 & 
  2.268$\pm$  0.112 \\
\aliii $\lambda$1854 & 
 1854.7184 & 
      6266.52 & 
        -204$\pm$           3 & 
        -406$\pm$          30 to +          77$\pm$          73 & 
  1.180$\pm$  0.135 \\
\feii $\lambda$2586 & 
 2586.6499 & 
      8741.48 & 
        -136$\pm$           2 & 
        -364$\pm$          40 to +          67$\pm$          17 & 
  1.584$\pm$  0.127 \\
\hline
\multicolumn{6}{c}{Reg2}\\ \hline
\sIiv $\lambda$1393 & 
 1393.7603 & 
      4708.78 & 
        -171$\pm$           5 & 
        -498$\pm$          46 to +         214$\pm$          33 & 
  1.811$\pm$  0.271 \\
\sIiv $\lambda$1402 & 
 1402.7729 & 
      4739.31 & 
        -166$\pm$          27 & 
        -412$\pm$          40 to +         175$\pm$          40 & 
  0.833$\pm$  0.233 \\
\sIii $\lambda$1526 & 
 1526.7070 & 
      5157.19 & 
        -214$\pm$           0 & 
        -654$\pm$          29 to +         245$\pm$          25 & 
  2.492$\pm$  0.203 \\
\civ $\lambda$1548 & 
 1548.2041 & 
      5232.25 & 
--- & 
--- & 
  4.533$\pm$  0.372$^f$ \\
\civ $\lambda$1550 & 
1550.7812 & 
      5232.25 & 
--- & 
--- & 
  4.533$\pm$  0.372$^f$ \\
\alii$\lambda$1670 & 
 1608.4510 & 
      5434.96 & 
        -124$\pm$          22 & 
        -396$\pm$          36 to +         159$\pm$          94 & 
  0.582$\pm$  0.151 \\
\feii $\lambda$1608 & 
 1670.7886 & 
      5643.60 & 
        -230$\pm$           5 & 
        -551$\pm$          44 to +         104$\pm$          58 & 
  1.626$\pm$  0.155 \\
\aliii $\lambda$1854 & 
 1854.7184 & 
      6265.91 & 
        -181$\pm$           6 & 
        -392$\pm$          54 to +          37$\pm$          54 & 
  1.295$\pm$  0.172 \\
\feii $\lambda$2586 & 
 2586.6499 & 
      8740.02 & 
        -134$\pm$           3 & 
        -321$\pm$          44 to +          75$\pm$          20 & 
  1.503$\pm$  0.169 \\
\hline
\multicolumn{6}{c}{Reg3}\\ \hline
\sIiv $\lambda$1393 & 
 1393.7603 & 
      4710.30 & 
         -79$\pm$           3 & 
        -421$\pm$          40 to +         223$\pm$          53 & 
  1.975$\pm$  0.304 \\
\sIiv $\lambda$1402 & 
 1402.7729 & 
      4741.52 & 
         -31$\pm$           7 & 
        -362$\pm$          13 to +         265$\pm$          33 & 
  1.607$\pm$  0.277 \\
\sIii $\lambda$1526 & 
 1526.7070 & 
      5157.64 & 
        -192$\pm$          24 & 
        -761$\pm$          25 to +         292$\pm$          25 & 
  2.773$\pm$  0.397 \\
\civ $\lambda$1548 & 
 1548.2041 & 
      5233.11 & 
--- & 
--- & 
  4.718$\pm$  0.345$^f$ \\
\civ $\lambda$1550 & 
1550.7812 & 
      5233.11 & 
--- & 
--- & 
  4.718$\pm$  0.345$^f$ \\
\alii$\lambda$1670 & 
 1608.4510 & 
      5436.28 & 
         -56$\pm$           7 & 
        -346$\pm$          40 to +         289$\pm$          36 & 
  1.388$\pm$  0.187 \\
\feii $\lambda$1608 & 
 1670.7886 & 
      5646.23 & 
         -95$\pm$           5 & 
        -348$\pm$          44 to +         134$\pm$          44 & 
  1.396$\pm$  0.161 \\
\aliii $\lambda$1854 & 
 1854.7184 & 
      6267.16 & 
        -125$\pm$           4 & 
        -353$\pm$          24 to +         136$\pm$          76 & 
  1.386$\pm$  0.152 \\
\feii $\lambda$2586 & 
 2586.6499 & 
      8741.82 & 
         -76$\pm$           6 & 
        -258$\pm$          53 to +          97$\pm$          24 & 
  1.896$\pm$  0.292 \\
\hline
\multicolumn{6}{c}{Reg4}\\ \hline
\sIiv $\lambda$1393 & 
 1393.7603 & 
      4711.37 & 
          34$\pm$          44 & 
        -200$\pm$          73 to +         256$\pm$           6 & 
  1.171$\pm$  0.624 \\
\sIiv $\lambda$1402 & 
 1402.7729 & 
      4738.45 & 
        -178$\pm$           8 & 
        -403$\pm$          36 to +          65$\pm$          61 & 
  1.335$\pm$  0.465 \\
\sIii $\lambda$1526 & 
 1526.7070 & 
      5158.26 & 
        -111$\pm$          29 & 
        -485$\pm$          97 to +         254$\pm$          33 & 
  2.495$\pm$  0.383 \\
\civ $\lambda$1548 & 
 1548.2041 & 
      5233.07 & 
--- & 
--- & 
  4.138$\pm$  0.553$^f$ \\
\civ $\lambda$1550 & 
1550.7812 & 
      5233.07 & 
--- & 
--- & 
  4.138$\pm$  0.553$^f$ \\
\alii$\lambda$1670 & 
 1608.4510 & 
      5436.88 & 
          22$\pm$           0 & 
        -165$\pm$          98 to +         273$\pm$         107 & 
  0.940$\pm$  0.431 \\
\feii $\lambda$1608 & 
 1670.7886 & 
      5645.66 & 
         -79$\pm$           7 & 
        -260$\pm$          34 to +         132$\pm$          41 & 
  1.330$\pm$  0.214 \\
\aliii $\lambda$1854 & 
 1854.7184 & 
      6266.40 & 
        -116$\pm$           4 & 
        -350$\pm$          46 to +         128$\pm$          29 & 
  1.804$\pm$  0.266 \\
\feii $\lambda$2586 & 
 2586.6499 & 
      8741.22 & 
         -51$\pm$           2 & 
        -185$\pm$          15 to +         108$\pm$          40 & 
  1.355$\pm$  0.234 \\
\hline

%% file: tables/logdensity.tex
Reg1 &    2.38100$\pm$   0.00012 &           84$\pm$          10 &      0.6$\pm$     0.1 &     30.6$\pm$     3.4 &       24.8$\pm$    3.8&    1.23$^{+   0.22}_{-   0.19}$ &    3.92$^{+   0.30}_{-   1.01}$ \\
Reg2 &    2.38041$\pm$   0.00014 &           75$\pm$          12 &      0.7$\pm$     0.1 &     51.0$\pm$     8.1 &       53.8$\pm$    8.9&    0.94$^{+   0.22}_{-   0.18}$ &    4.36$^{+   0.21}_{-   0.32}$ \\
Reg3 &    2.38046$\pm$   0.00012 &           60$\pm$           8 &      0.8$\pm$     0.1 &     39.8$\pm$     5.2 &       38.8$\pm$    6.3&    1.03$^{+   0.21}_{-   0.17}$ &    4.24$^{+   0.21}_{-   0.34}$ \\
Reg4 &    2.37994$\pm$   0.00027 &           83$\pm$          23 &      1.2$\pm$     0.2 &     36.6$\pm$     7.8 &       33.5$\pm$    8.2&    1.06$^{+   0.39}_{-   0.24}$ &    4.20$^{+   0.31}_{-   0.73}$ \\

%% file: tables/SFR_tab.tex
Reg1 &  0.35$\pm$ 0.14 & 
  84.6$\pm$   3.3 & 
   9.7$\pm$   5.5 & 
  1.3$\pm$  0.7 & 
          10$\pm$           6 & 
  0.1$\pm$  0.6\\
Reg2 &  0.30$\pm$ 0.07 & 
 155.0$\pm$   8.7 & 
  14.4$\pm$   4.6 & 
  1.9$\pm$  0.6 & 
          15$\pm$           5 & 
  0.5$\pm$  0.3\\
Reg3 &  0.50$\pm$ 0.08 & 
  72.0$\pm$   5.6 & 
  15.7$\pm$   5.7 & 
  2.0$\pm$  0.7 & 
          16$\pm$           6 & 
  0.4$\pm$  0.4\\
Reg4 &  0.43$\pm$ 0.03 & 
  48.1$\pm$   7.2 & 
   7.7$\pm$   1.4 & 
  1.0$\pm$  0.2 & 
           8$\pm$           1 & 
  0.3$\pm$  0.2\\